\documentclass[12pt]{article}
\pdfoutput=1
\usepackage{geometry}
\usepackage{amsmath}
\usepackage{amssymb}
\usepackage{graphicx}
\usepackage{esint}

\usepackage{ifpdf}
\usepackage{hyperref}
\usepackage{epsfig}
\usepackage{amsfonts}
\usepackage{amsmath}
\usepackage{amssymb}
\usepackage{color} 
\usepackage{graphicx}
\usepackage{grffile}
\usepackage{subfig}
\usepackage{notoccite}

\newcommand{\be}{\begin{equation}}
\newcommand{\ee}{\end{equation}}
\newcommand{\ba}{\begin{eqnarray}}
\newcommand{\ea}{\end{eqnarray}}
\newcommand{\bi}{\begin{itemize}}
\newcommand{\ei}{\end{itemize}}

\newcommand{\aslash}[1]{\,\,{\raise.15ex\hbox{/}\mkern-12mu #1}}
\newcommand{\bslash}[1]{\,\,{\raise.15ex\hbox{/}\mkern-9mu #1}}

\renewcommand{\bar}{\overline}
\renewcommand{\tilde}{\widetilde}
\renewcommand{\hat}{\widehat}

\begin{document}

\begin{titlepage}

\begin{center}
\vspace{1cm}
{\Large \bf  Polarised Black Holes in AdS}
\vspace{0.8cm}

{\large  Miguel S. Costa$^{\dagger,\ddagger}$, Lauren Greenspan$^\dagger$, Miguel Oliveira$^\dagger$,  \\
Jo\~ao Penedones$^{\dagger,\ddagger}$, Jorge E. Santos$ ^{\Diamond}$}

\vspace{.5cm}

{\it  $ ^\dagger$Centro de F\'\i sica do Porto,
Departamento de F\'\i sica e Astronomia\\
Faculdade de Ci\^encias da Universidade do Porto\\
Rua do Campo Alegre 687,
4169--007 Porto, Portugal}
\\
\vspace{.3cm}
{\it  $ ^\ddagger$Theory Division, Department of Physics, CERN\\
CH-1211 Gen\`eve 23, Switzerland}
\\
\vspace{.3cm}
{\it  $ ^\Diamond$Department of Applied Mathematics and Theoretical Physics\\
University of Cambridge, Wilberforce Road\\
Cambridge CB3 0WA, UK}

\end{center}
\vspace{1cm}

\begin{abstract}
We consider solutions in Einstein-Maxwell theory with a negative cosmological constant that  asymptote to global $AdS_{4}$ with conformal boundary $S^{2}\times\mathbb{R}_{t}$. At the  sphere at infinity we turn on a space-dependent electrostatic potential, which does not destroy the  asymptotic $AdS$ behaviour. For simplicity we focus on the case of a dipolar electrostatic potential. We find two new geometries: (i) an $AdS$ soliton that includes the full backreaction of the electric field on the $AdS$ geometry; (ii) a polarised neutral black hole that is deformed by the electric field, accumulating  opposite charges in each hemisphere. For both geometries  we study boundary data such as the charge density and the stress tensor. For the black hole we also study the horizon charge density and area, and further verify a Smarr formula. Then we consider this system at  finite temperature and compute the Gibbs free energy for both  $AdS$ soliton and black hole phases. The corresponding phase diagram generalizes the Hawking-Page phase transition. The $AdS$ soliton dominates the low temperature phase and the black hole the high temperature phase, with a  critical temperature that decreases  as the external electric field increases. Finally, we consider the simple case of a free charged scalar field on $S^{2}\times\mathbb{R}_{t}$ with conformal coupling. For a field 
in the $SU(N)$ adjoint representation we compare the phase diagram with the above gravitational system.
\end{abstract}

\bigskip
\bigskip

\end{titlepage}

\section{Introduction}

What happens if we place a neutral black hole in a background electric field? Intuitively, one expects that the BH should polarize and deform in response to the applied electric field. We will show that this intuition is indeed correct but, in order to be rigorous, we first need to understand what it means to apply an electric field in general relativity. Throughout this paper, we will restrict ourselves to four-dimensional spacetimes.

Let us start by considering classical solutions of Einstein-Maxwell theory
\be
R_{\mu\nu}=2F_{\mu\alpha}F_{\nu}^{\ \alpha}-\frac{1}{2}g_{\mu\nu}F_{\alpha\beta}F^{\alpha\beta}\,,\qquad\nabla_{\mu}F^{\mu\nu}=0\,,
\ee
with $F =  d {\cal A}$. This is the minimal theory that can contain polarised black holes. Remarkably, 40 years ago Ernst \cite{Ernst} constructed the following exact solution 
\be
 ds^{2}=\lambda^{2}(y,\theta)\left[-\left(1-\frac{y_{0}}{y}\right)dt^{2}+\frac{dy^{2}}{1-\frac{y_{0}}{y}}+y^{2}d\theta^{2}\right]+\frac{y^{2}\sin^{2}\theta d\phi^{2}}{\lambda^{2}(y,\theta)}\,
,\qquad {\cal A}={\cal E}(y_{0}-y)\cos\theta dt\,,
\label{ErnstSolution}
\ee
where $\lambda(y,\theta)=1+ ({\cal E}/2)^{2}y^{2}\sin^{2}\theta$. The parameter $y_{0}$ is the areal radius of the black hole and  ${\cal E}$ parameterizes the background electric field. In the absence of the black hole ($y_{0}=0$) and to linear order in ${\cal E}$ the solution describes flat spacetime with a constant electric field along $z=y\cos\theta$.

This solution realizes the intuition of black hole polarization and deformation. The charge density on the horizon (or electric flux) is given by 
\be
\frac{ dQ}{ d\Omega}=\frac{1}{4\pi}\frac{2y_{0}\gamma\cos\theta}{\left(1+\gamma^{2}\sin^{2}\theta\right)^2}\,,
\qquad\gamma=\frac{{\cal E}y_{0}}{2}\,.
\ee
As expected, the electric field ${\cal E}$ induces a positive charge density in the upper hemisphere $0\le\theta<\frac{\pi}{2}$ and a negative charge density in the lower hemisphere $\frac{\pi}{2}<\theta\le\pi$. The horizon also gets deformed. The equatorial perimeter shrinks to $2\pi y_{0} /(1+\gamma^{2})$ while the length of a meridian expands to $\pi y_{0}\left(1+\gamma^{2}/2\right)$. While our intuition works at small distances, the backreaction of the electric field drastically changes the geometry far from the black hole. To see that consider the length of an equatorial circle ($\theta=\pi/2$) as a function of the coordinate $y$ in the solution without black hole ($y_{0}=0$). This length increases from zero at $y=0$ until it reaches a maximum at $y=2/{\cal E}$, and then decreases towards zero as $y\to\infty$. This shows that our original picture of a black hole placed in an approximately flat spacetime with a background electric field is only realized for $y_{0}\ll 2/{\cal E}$ or $\gamma\ll1$. For strong electric fields, i.e. $\gamma\gg1$, the Ernst solution describes different physics.

In this paper we study polarised black holes in Anti-de Sitter ($AdS$) spacetime. One motivation is the study of neutral black hole polarization in a context where the asymptotic geometry of spacetime is not destroyed by the presence of a background electric field. Another motivation is the study of conformal theories subject to an external electric field.
We consider the action
\be
S=\frac{1}{16\pi G_N} \int d^4x \sqrt{-g} \left( R  + \frac{6}{l^2}  -F_{\alpha\beta}F^{\alpha\beta} \right) + \frac{1}{8\pi G_N} \int d^3x \sqrt{h} K
\ee
where $l$ is the $AdS$ length scale and we also added the Gibbons-Hawking-York boundary term. The 
field equations are
\be
R_{\mu\nu}+\frac{3}{l^2}\,g_{\mu\nu}=2F_{\mu\alpha}F_{\nu}^{\ \alpha}-\frac{1}{2}g_{\mu\nu}F_{\alpha\beta}F^{\alpha\beta}\,,
\qquad\nabla_{\mu}F^{\mu\nu}=0\,.
\ee 
We will look for solutions which asymptote to global $AdS_{4}$ with a conformal boundary given by $S^{2}\times\mathbb{R}_{t}$. In the gauge/gravity duality a $U(1)$ gauge field 
${\cal A}_\alpha$ in the bulk is dual to a global current operator $J_a$ in the boundary theory, with lower case latin indices running over the boundary coordinates. In general, turning on a source $C_a$ for  the operator $J_a$ on the boundary theory corresponds to a non-normalizable mode of the bulk gauge field. For  a source given by a generic electrostatic potential 
\be
C_t=\Phi(\theta,\phi) = \sum_{l,m} a_{l,m} Y_l^m(\theta,\phi)\,,
\ee
where $ Y_l^m$ are the usual scalar harmonics on $S^{2}$, the gauge field near the boundary will have the asymptotic behaviour
\be
{\cal A} \approx \big( \Phi(\theta,\phi) + 
4\pi G_N\rho(\theta,\phi)\,z \big) dt \,,
\label{eq:A_asympt}
\ee
where $\{t,\theta,\phi,z\}$ are Fefferman-Graham coordinates in $AdS$ \cite{Fefferman:2007rka}. This means that we can turn on any space dependent electrostatic potential (or chemical potential) on the sphere at infinity, without destroying the {\em good} asymptotic $AdS$ behaviour. In other words, the dual theory may be deformed by the relevant operator $C_a J^a$ without altering the UV physics. The response function $\rho(\theta,\phi)$ reads the charge density, that is, how charges are distributed on the sphere due to the interaction with the electric field. 

\begin{figure}[t!]
\centering
\includegraphics[width=100mm]{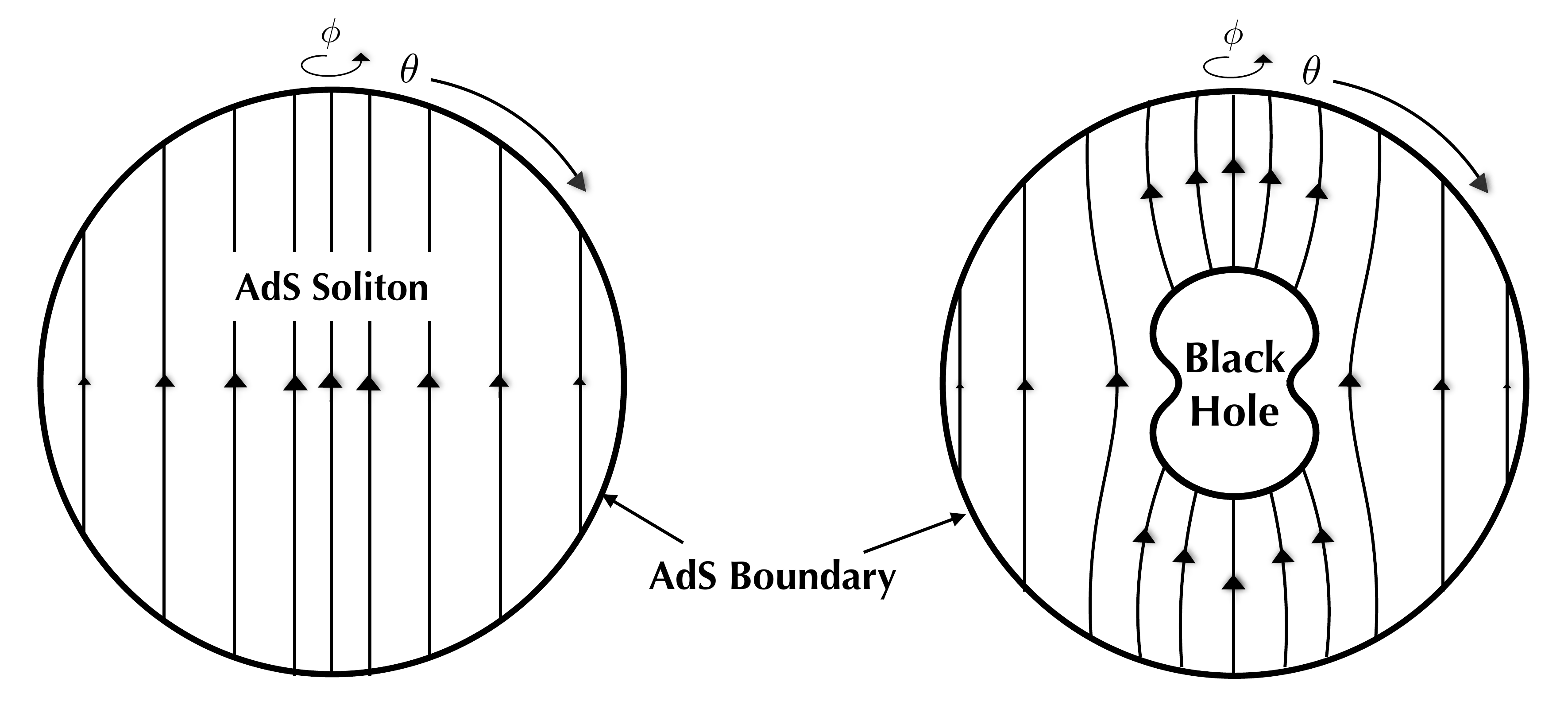}
\caption{Pictorial representation of the two new geometries found in this work. (Left) $AdS$ soliton with the electric field throughout space; 
(Right) Black hole polarised by  the electric field. 
\label{Soliton&BH}}
\end{figure}

For simplicity in this paper we shall consider the particular case of a dipolar potential
\be
\Phi(\theta)= {\mathcal {\cal E}}\cos\theta 
\label{eq:DipolarSource}
\ee
at the conformal boundary. With this boundary condition we will construct horizonless solutions describing an $AdS$ soliton with a non-trivial electric field.  
It describes the vacuum of the dual deformed CFT.
Then we construct  polarised black hole solutions with the same boundary conditions.
Figure \ref{Soliton&BH} gives a pictorial description of both geometries. We also study the phase diagram of this system at finite temperature, generalizing the Hawking-Page phase transition for the case of an external electric field. A simple analysis of a free charged scalar field  on $S^{2}\times\mathbb{R}_{t}$, with conformal coupling, shows a qualitatively similar phase diagram. In an appendix we present the perturbative analysis of the $AdS$ soliton solution up to third order in ${\cal E}$, which provides a test of our numerics.\footnote{Preliminary reports of the findings presented in this paper appeared in \cite{MiguelMSc, MiguelEssay}.}

\section{Electric flux in AdS} \label{sec:soliton}

Let us first consider the new $AdS$ soliton that results from turning on an electric field. In the case of a geometry with axial symmetry  
a convenient ansatz for our numerical implementation is (setting the $AdS$ radius $l=1$)
\begin{align}
ds^{2}=&\,\frac{1}{(1-r^{2})^{2}}
\Bigg\{ A(r,x)\, d\tau^{2}+\frac{4\,G(r,x)  dr^2}{2-r^2} 
\nonumber\\
&\qquad \qquad \quad  + r^2(2-r^2)\left[\frac{4\,C(r,x)}{2-x^2}\,\left( dx+\frac{H(r,x)  dr}{r}\right)^2+B(r,x)\,(1-x^2)^2  d\phi^2\right]\Bigg\} \,,
\label{AdSansatz}\\
{\cal A}=&-i\,r\,D(r,x)\, d\tau\,,
\nonumber
\end{align}
where $r$ is a radial coordinate running
from $r=0$ at the center of space,  to $r=1$ at the $AdS$ boundary. We shall work in 
the Euclidean setting with the time coordinate $\tau$ periodically identified and defined by $t=-i\tau$, as usual.
Global $AdS$ corresponds to $A=G=B=C=1$ and $H=D=0$. In this case, the usual radial $AdS$ coordinate $y$ is related to $r$ by $y=r\sqrt{2-r^2}/(1-r^{2})$ and $x$ is related to the usual polar angle on the S$^2$ via $x\sqrt{2-x^2}\equiv \cos\theta$.  $x\in [-1,1]$ parametrises the polar coordinate on $S^2$. Since we consider a dipolar source (\ref{eq:DipolarSource}), there is a reflection $x\to -x$ symmetry (corresponding in the usual polar coordinates to $\theta \to \pi-\theta$) so we can restrict the domain  to $x\in [0,1]$. We will use the reflection properties of the functions $A$, $G$, $C$, $B$, $H$, and $D$ to discretise the equations of motion on a domain bounded by $r=0, 1$ and $x=0,1$. Since the point $r=0$ is not a boundary but a fixed point of the coordinate system, we need to impose that the geometry is smooth at this point. 
For the above ansatz this is the same as imposing that the first derivatives along $r$ vanish at $r=0$. The same logic applies to the fixed point at the axis of symmetry $x=1$.  In this case smoothness implies that the first derivatives along $x$ vanish at $x=1$, together with $B(r,1)=C(r,1)$ and $H(r,1)=0$. 
Finally at $x=0$, the fixed point of the symmetry $x\to - x$, we require that the first derivatives along $x$ vanish, except for the functions $D(r,x)$ and $H(r,x)$ which vanish at that point. 

In the end we are left with one real boundary at $r=1$. Here we impose that the metric approaches that of global $AdS$ by setting
\be
A(1,x)=B(1,x)=C(1,x)=G(1,x)=1\,, \ \ \  H(1,x)=0\,.
\ee
We turn on the gauge field by using the boundary condition
\be
D(1,x)=\Phi(x)\,,
\ee
where $\Phi(x)={\mathcal {\cal E}}x\sqrt{2-x^2}$ is the dipolar potential (\ref{eq:DipolarSource}) expressed in terms of $x$.

We find the solutions using the  Einstein-deTurck equations with reference metric given by the functions $A(r,x)=B(r,x)=C(r,x)=G(r,x)=1$ and $H(r,x)=0$ in (\ref{AdSansatz}), which gives a set of elliptic PDE's for the six functions of (\ref{AdSansatz}). We start with pure $AdS$ as a seed for a solution with small electric field parameter ${\cal E}$, and then increase ${\cal E}$ using each solution as a seed for the next. For a review of these methods see \cite{Dias:2015nua}.

\subsection{Results}

\begin{figure}[t!]
\centering
\subfloat[ ]{
\includegraphics[width=50mm]{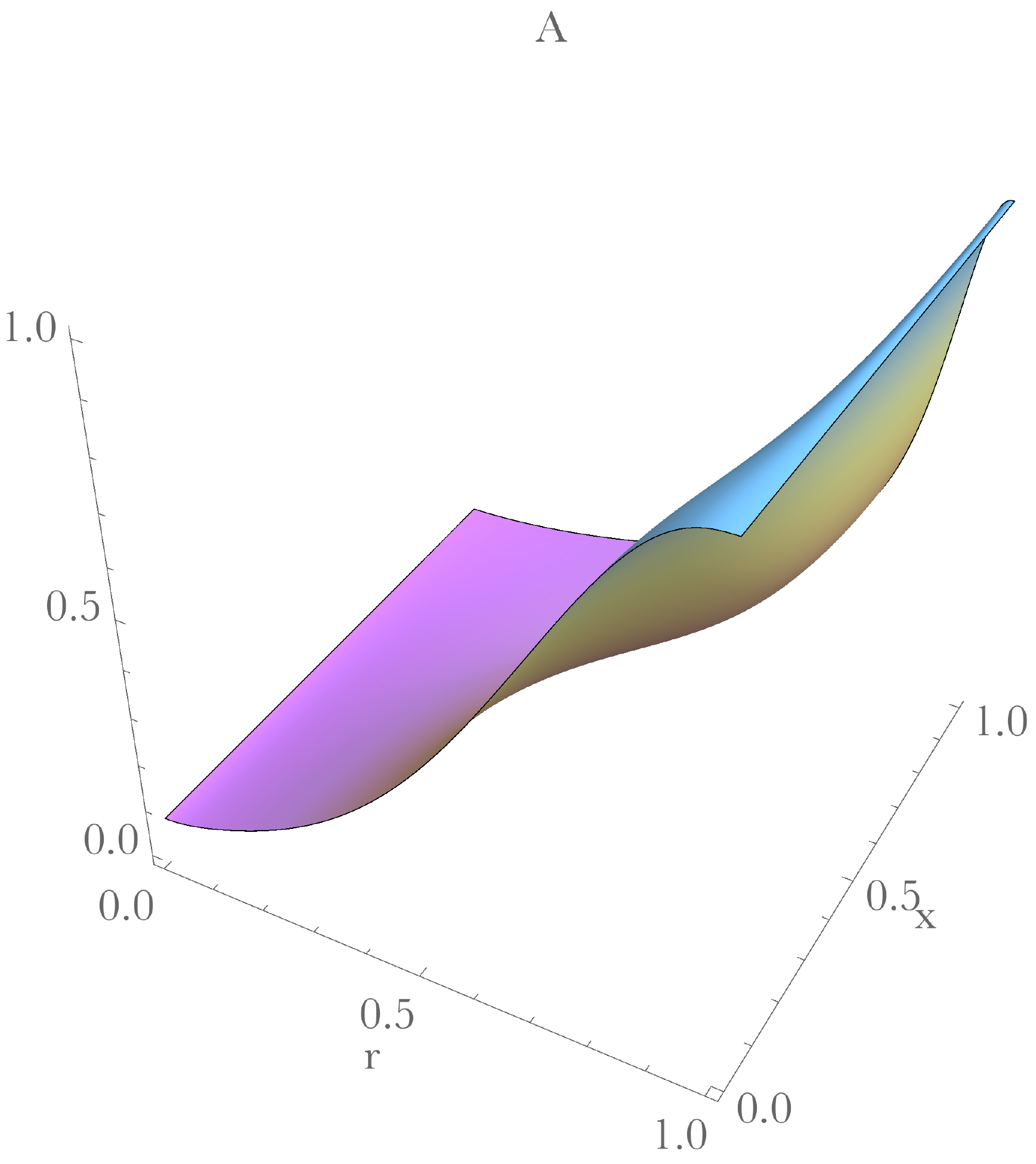}
}
\subfloat[ ]{
\includegraphics[width=50mm]{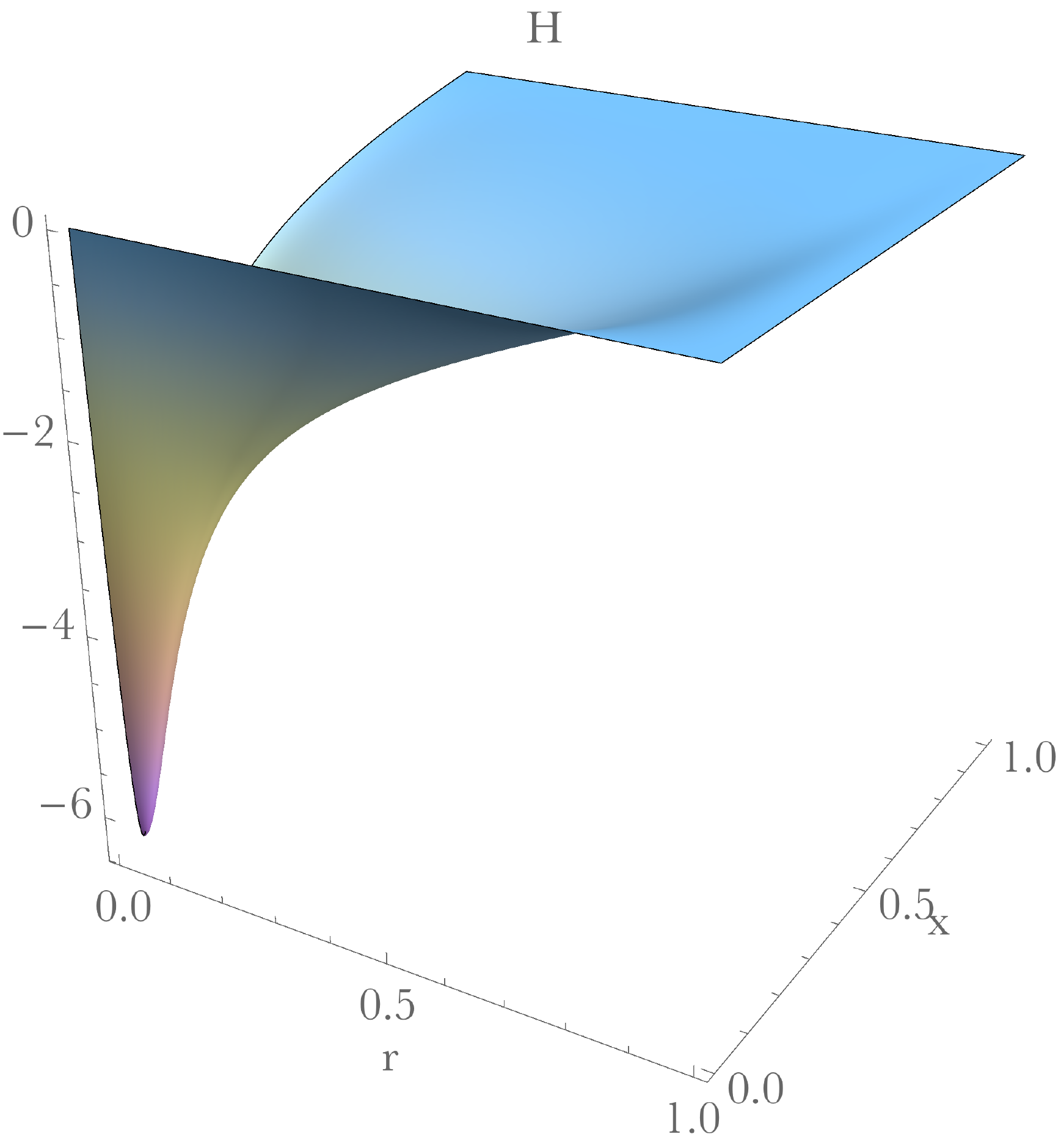}
}
\subfloat[ ]{
\includegraphics[width=50mm]{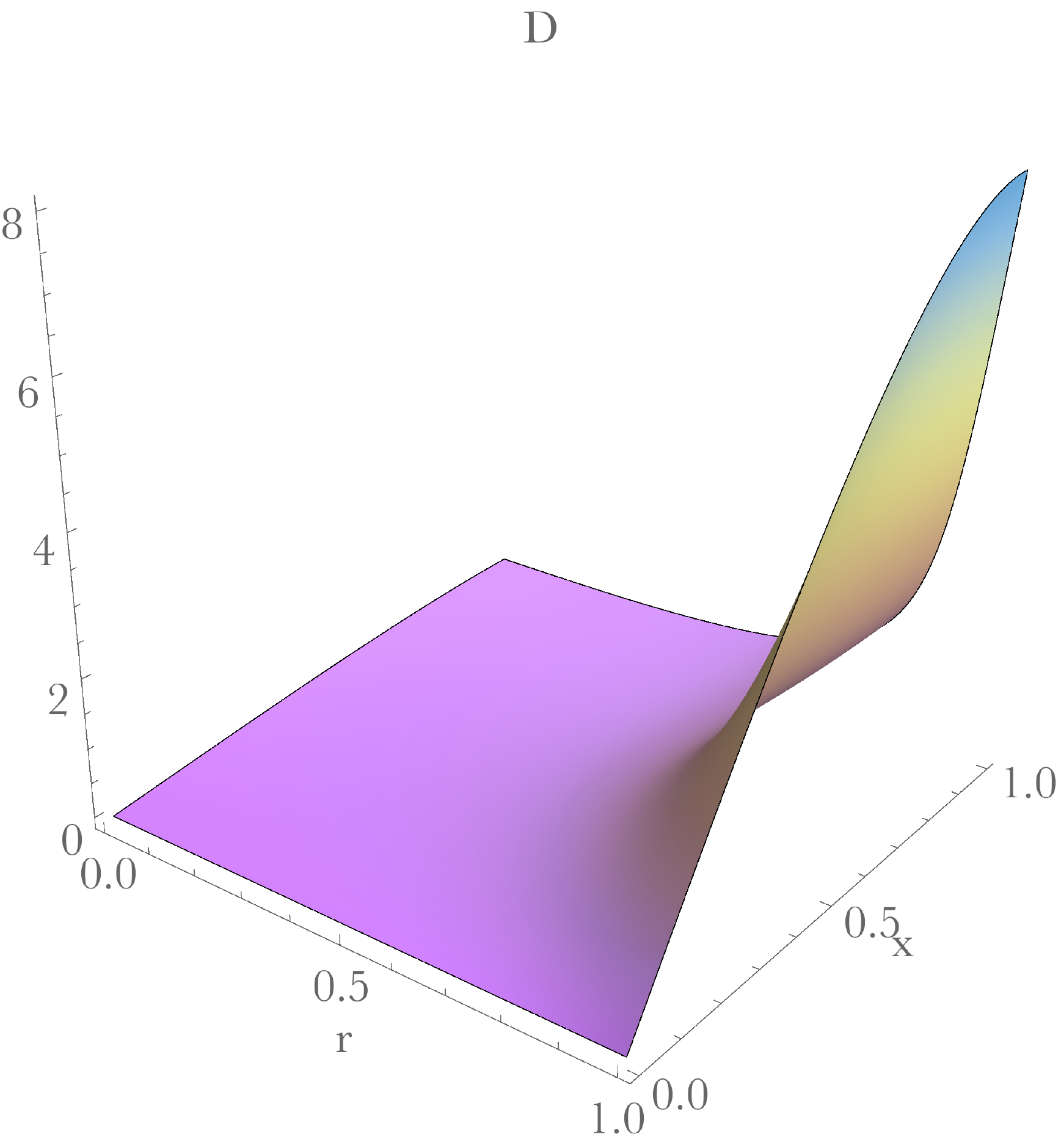}
}
\caption{Functions $A$, $H$ and $D$ for ${\cal E}=8.028$. In this case we used a numerical grid of $200\times 200$ points.
\label{metnoBH}}
\end{figure}

\begin{figure}[t!]
\centering
\includegraphics[width=65mm]{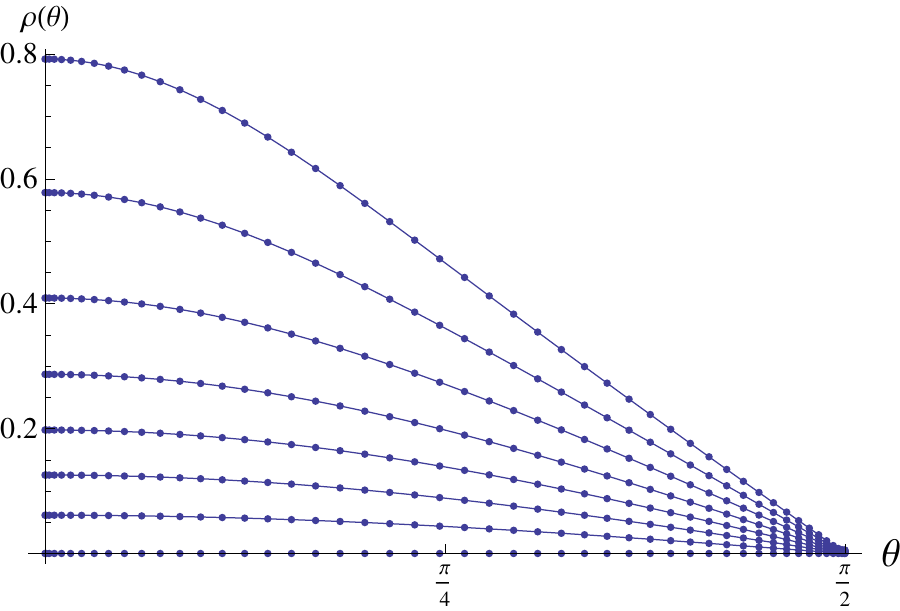}
\caption{The charge density on the $AdS$ boundary for several values of ${\cal E} \in [0,4.4] $ with $G_N$ set to one
\label{chargedensAdS}}
\end{figure}

In figure \ref{metnoBH} we show some of the functions in the ansatz (\ref{AdSansatz}) for a value of the electric field ${\cal E}$. We can now calculate several boundary observables, including the charge density $\rho(\theta)$ defined in (\ref{eq:A_asympt}), which can be expressed as 
\be
\frac{1}{4\pi G_N}\star \left.  \!F\right|_{r=1}=\rho(\theta)\, d\Omega_{2}\,,
\label{eq:carge_dens}
\ee
where $ d\Omega_2$ is the volume form on the unit $S^2$. This is plotted in figure \ref{chargedensAdS} for several values of the electric field magnitude. The charge density is maximal at the pole and vanishes at the equator, as expected from the choice of boundary condition.

\begin{figure}[t!]
\centering
\subfloat[]{
\includegraphics[height=35mm]{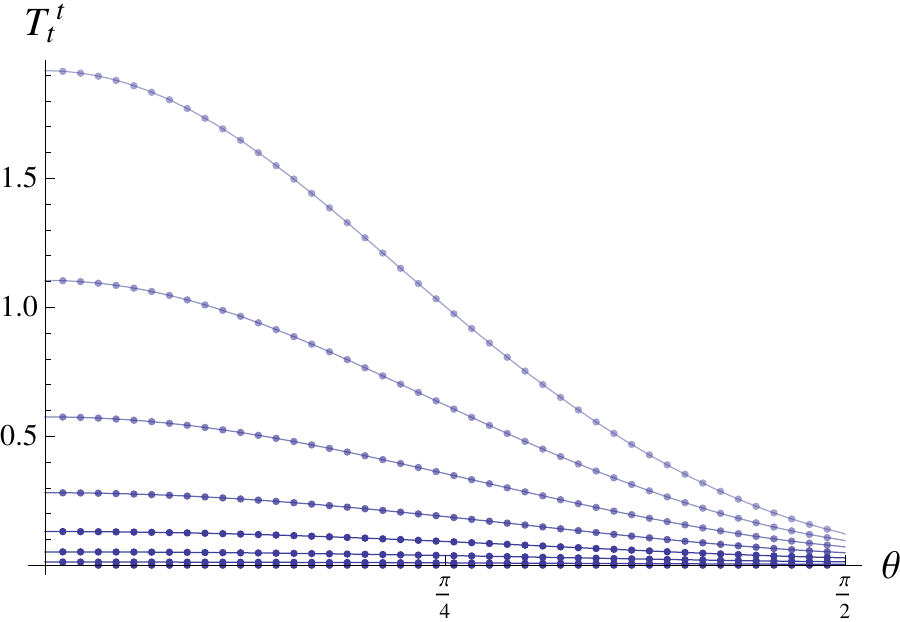}
\label{endensAdS}
}
\subfloat[]{
\includegraphics[height=35mm]{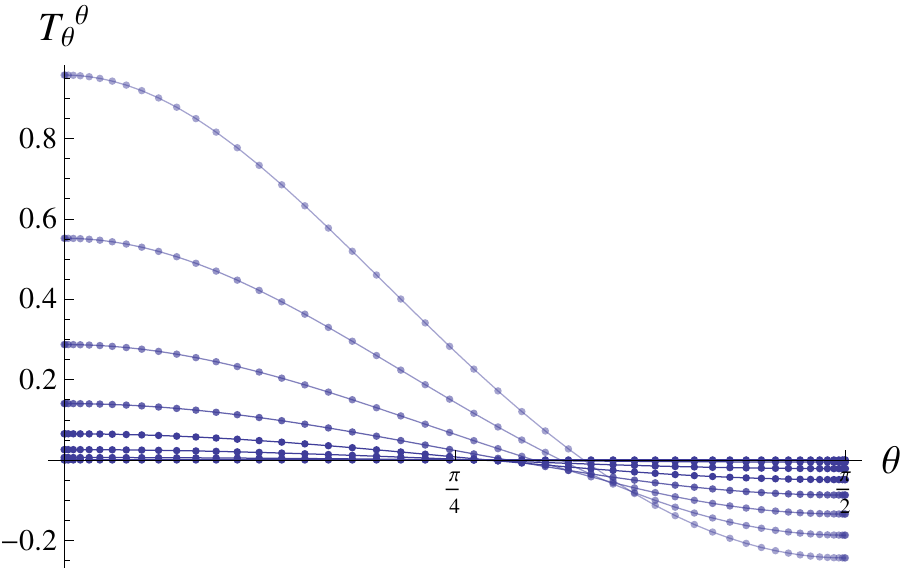}
\label{TththAdS}
}
\subfloat[]{
\includegraphics[height=35mm]{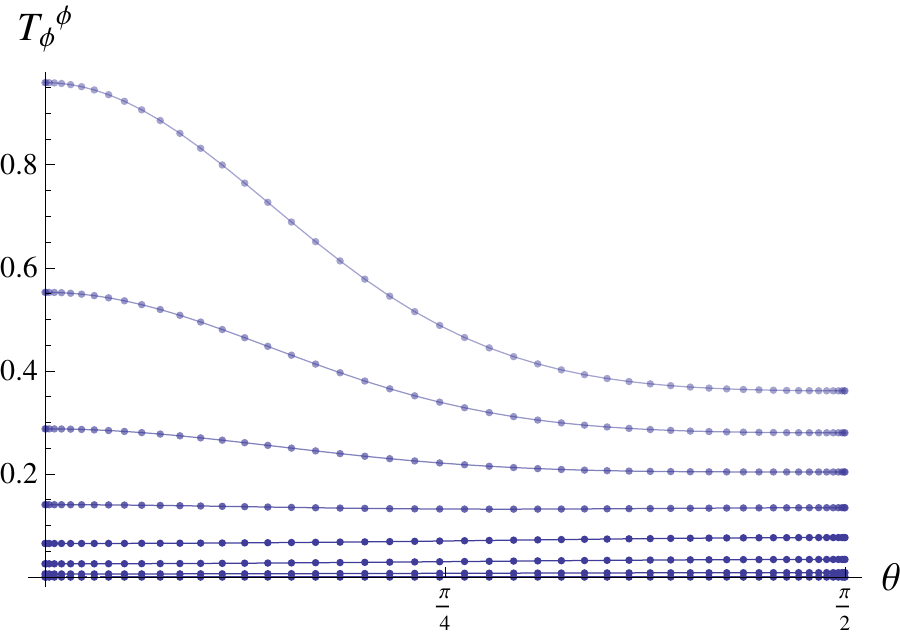}
\label{TffAdS}
}
\caption{(a) Energy density on the $AdS$ boundary for several values of ${\cal E}$;
(b)-(c) Spatial components of the boundary stress tensor for several values of ${\cal E}\in [0,4.4]$.
In these plots we set $G_N=1$.}
\end{figure}

\begin{figure}[b!]
\centering
\includegraphics[width=60mm]{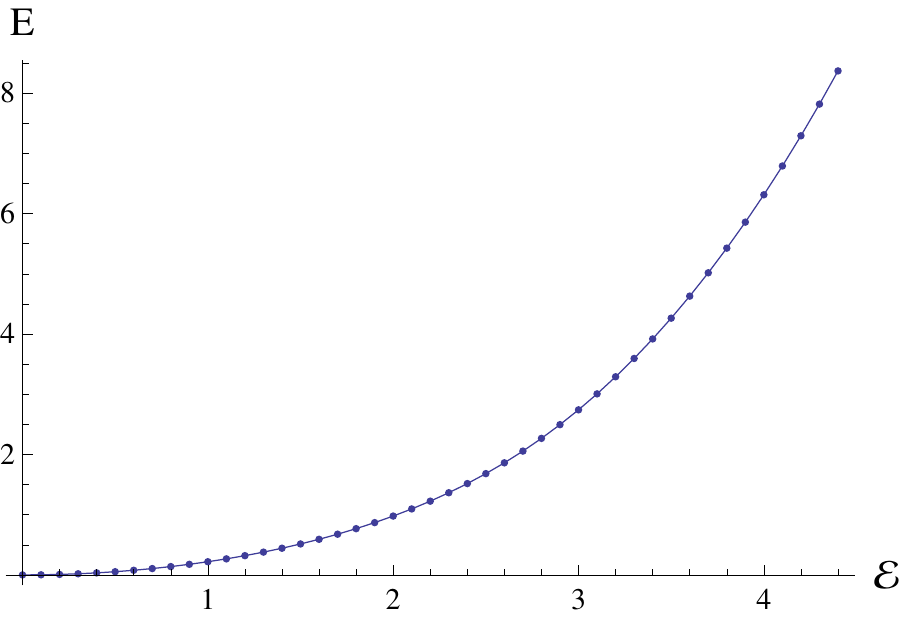}

\caption{Energy of the 
boundary theory as a function of the electric field ${\cal E}$ (setting $G_N=1$).\label{energyAdS}}
\end{figure}

We can also compute the stress tensor of the boundary theory \cite{gr-qc/9209012,Skenderis:2002wp}. Because we are working in $AdS$, the action diverges due to a cosmological constant term which is proportional to the volume of spacetime. We must cancel this divergence by adding to it a counter term
of the form  \cite{hep-th/9902121,deHaro:2000xn} 
\be
S_{CT}=-\frac{1}{8\pi G_N}\int_{\partial M} d^{3}x\sqrt{h}\left(1-\frac{l^{2}}{12}\,R\right),
\ee
where $h$ is the induced metric on the boundary. We can then derive the stress tensor from the on-shell action
(setting $l=1$)
\be
T_{\mu\nu}=\frac{2}{\sqrt{h}}\frac{\delta S}{\delta h^{\mu\nu}}=
\frac{1}{8\pi G_N}  \left(K_{\mu\nu}-K\, h_{\mu\nu}+G_{\mu\nu}-2h_{\mu\nu}\right) .
\ee
The first two terms are the extrinsic curvature and its trace, respectively, and come from the Gibbons-Hawking-York boundary term in the on-shell action. The last two terms are from the counter term contribution, with $G_{\mu\nu}$ the Einstein tensor on the boundary. The stress tensor is evaluated using the asymptotic expansion of the metric functions
up to $O(1-r)^{5}$ including logs, for example,
\be
A(r,x)=\underset{i=0}{\sum}(1-r)^{i}\alpha_{i}(x) +\log(1-r) \,\underset{i=4}{\sum} (1-r)^{i}a_{i}(x)\,.
\ee
The $\alpha$'s that are not fixed by the equations of motion describe normalizable modes that are then fixed by the boundary conditions. We can determine these by computing derivatives of the appropriate numerical solutions and evaluating them at $r=1$. For the $AdS$ ansatz written above, the energy density only depends explicitly on the $g_{\tau\tau}$ metric component, that is on the function $A(r,x)$,
\be
T_{t}^{\,t}=-\frac{3\alpha_{3}(\theta)}{128\pi G_{N}}\,.
\ee
This is plotted in figure \ref{endensAdS}. Note that even though we used the coordinate $x$ in the numerics, we decided to plot all our boundary quantities as a function of $\theta$, since this is a more familiar coordinate on the $S^2$. Like the charge density, $T_{t}^{\,t}$ is maximal at the pole,  minimal at the equator and increases for increasing ${\cal E}$. 
In figure \ref{energyAdS} we also show the total energy of the boundary theory as a function of ${\cal E}$.

For the  spatial components of the stress tensor we have
\ba
T_{\theta}^{\,\theta}&=&\frac{-3\big(\alpha_{3}(\theta)+\text{\ensuremath{\beta_{3}}}(\theta)\big)}{128\pi G_{N}}\,,\\
 T_{\phi}^{\,\phi}&=&\frac{3\text{\ensuremath{\beta_{3}}}(\theta)}{128\pi G_{N}}\,.\notag
\ea
Here, $\beta_3$ is the third-order power-law mode for the metric function $B$. These are also plotted in figures \ref{TththAdS} and \ref{TffAdS}. The $\theta \theta$ component is positive for points below a critical value of $\theta$ dependent on ${\cal E}$, and negative otherwise. That is, near the equator
the fluid has negative pressure. The $\phi \phi$ component shows that the pressure along the $\phi$ direction decreases from the poles to the equator and is independent of $\phi$ as it should be since there is no net flow of momenta in that direction. All other components of the stress tensor vanish. The non-zero components can be related by an equation describing the conservation of energy and momentum 
in the presence of a background electric field
\be
\nabla_a T^{ab} + j_a F^{ab}=0\,,
\ee
where $j^a=(\rho,j^i)$ and $j^i$ is the current density on the sphere at the boundary. This equation arises as a consequence of the Ward identities. 
The only nontrivial component for our ansatz corresponds to $b=\theta$ and leads to the relation
\be
\partial_\theta \left( \sin \theta \,T^{\,\theta}_\theta \right) -\cos \theta \, T^{\,\phi}_\phi=-{\cal E}\rho(\theta) \sin^2 \theta\,.
\label{conservationT}
\ee
For ${\cal{E}} \neq 0$, this is obeyed by our numerical solutions with a precision of $10^{-8}$ relative to $T^{\,\phi}_\phi$.

\begin{figure}[t!]
\centering
\subfloat[]{
\includegraphics[width=65mm]{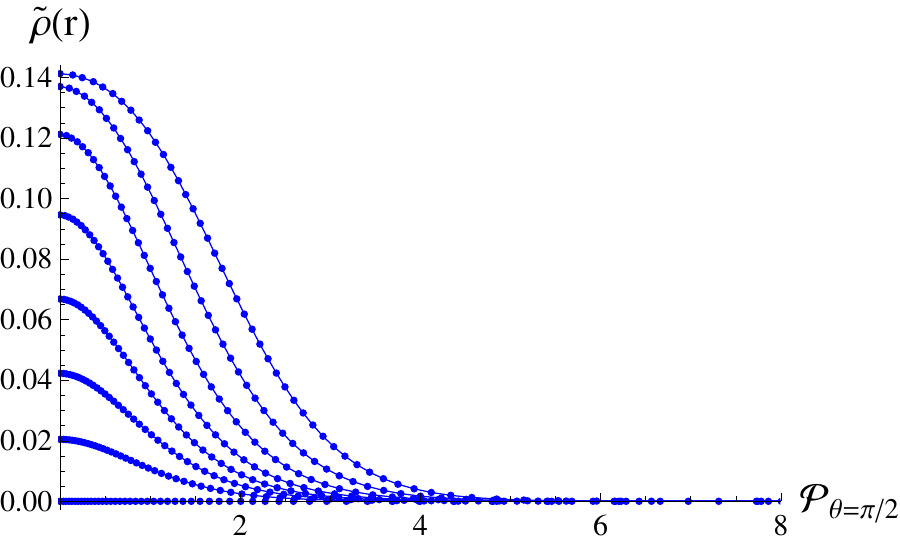}
}
\subfloat[]{
\includegraphics[width=60mm]{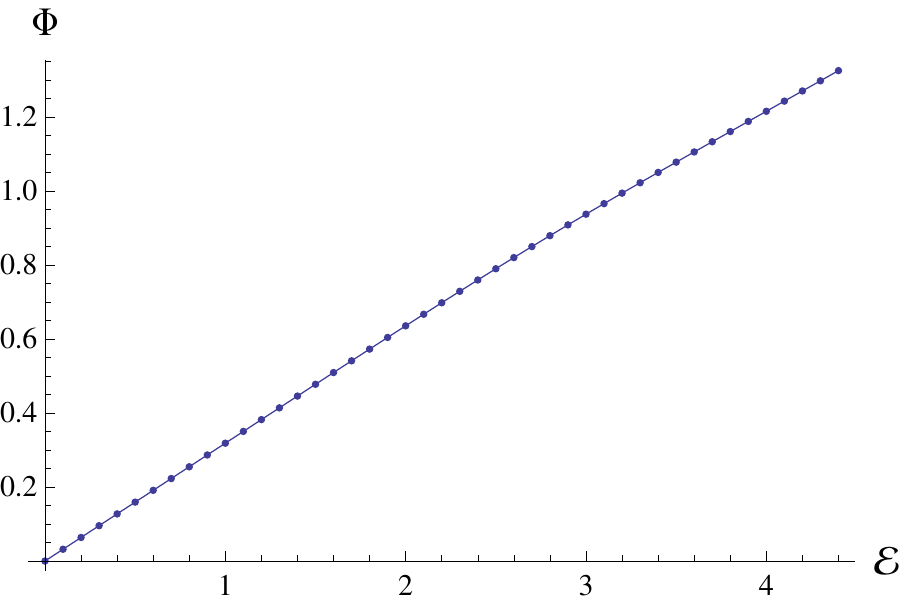}
}
\caption{(a) Flux density through the equatorial plan for several values of ${\cal E}$  as a function of the proper distance from the center; (b) The total flux through the equator as a function of ${\cal E}$
($G_N=1$).
\label{fluxAdS}}
\end{figure}

\begin{figure}[t!]
\centering
\subfloat[]{
\includegraphics[height=63mm]{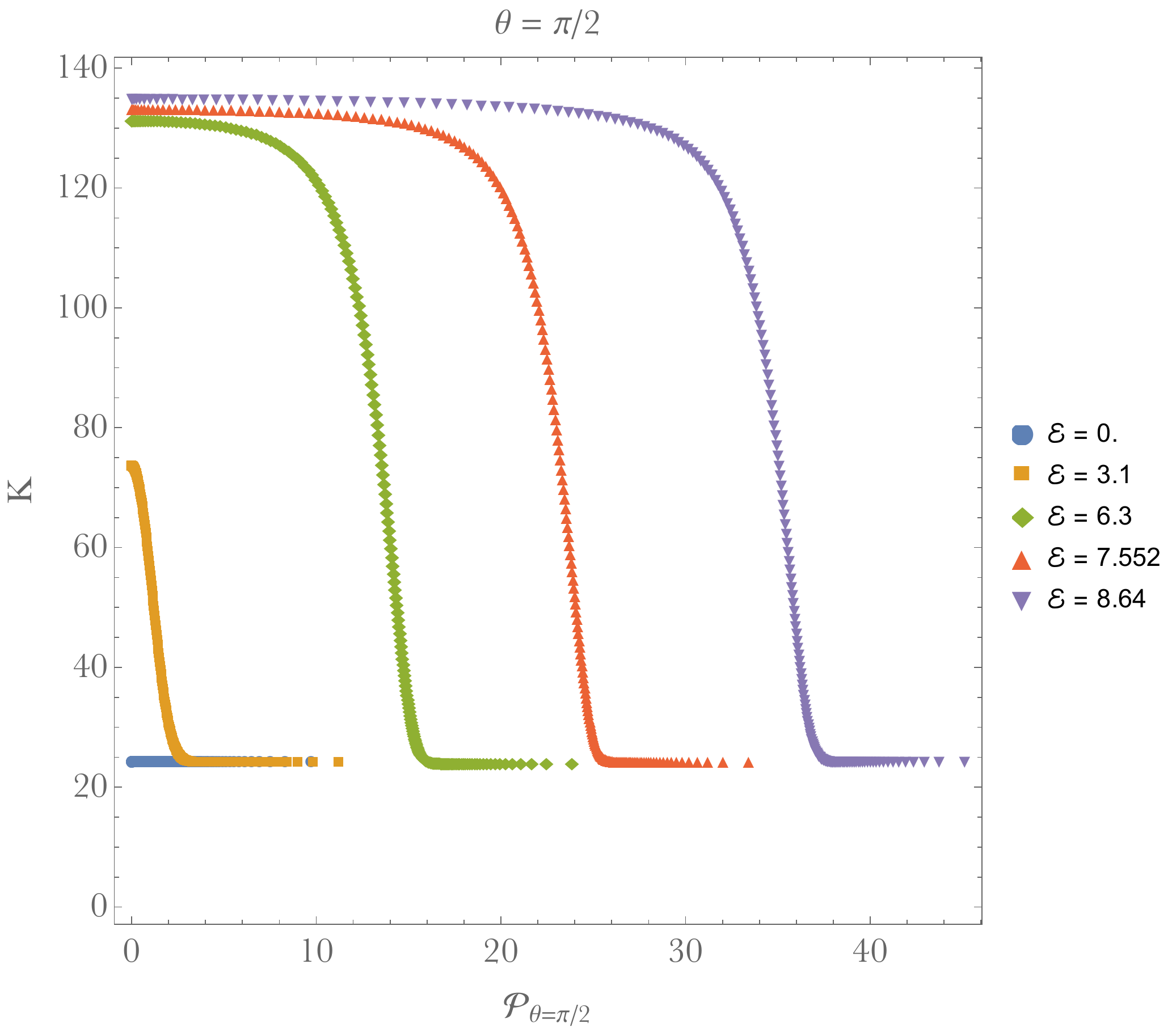}
}
\subfloat[]{
\includegraphics[height=60mm]{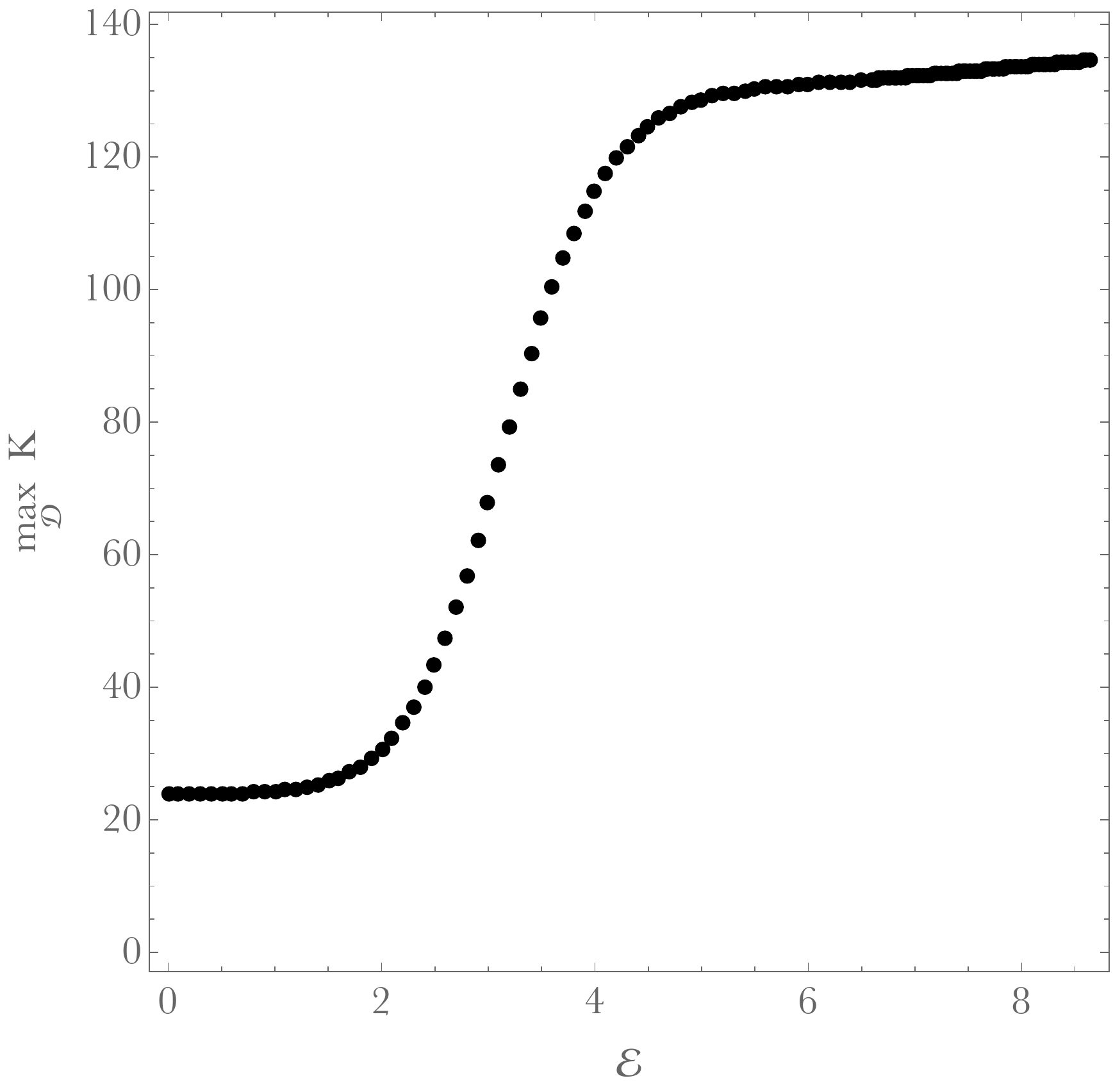}
}
\caption{(a)  Kretschman scalar invariant at the equator for several values of ${\cal E}$ as a function of the proper distance from the center; (b) maximum value of the Kretschman plotted with ${\cal E}$.
\label{kretsch}}
\end{figure}

Let us now  develop some intuition on the behaviour of the $AdS$ soliton in the bulk. We may consider the flux density $\tilde{\rho}$ through the $\theta=\pi/2$ plane, defined by
\be
\frac{1}{4\pi G_N}\star \left. \!F \right |_{\theta=\pi/2}=\tilde{\rho}(r)\sqrt{g_{rr}g_{\phi\phi}}\, dr\wedge  d\phi\,.
\ee
The flux density is zero at $r=1$ and maximal at the $AdS$ center. Figure \ref{fluxAdS}a plots this flux density in terms of the proper radial distance from the center of space along the equatorial plane. 
For a general direction this proper distance is given by
\be
{\cal P}_\theta = \int_0^r  \sqrt{g_{rr} (r',\theta)} \,dr'\,.
\ee
As the electric field ${\cal E}$ increases we see that the flux is more spread over space. Also, the total value of the flux increases with ${\cal E}$, as shown in figure \ref{fluxAdS}b. We may also consider the behaviour of the curvature throughout the space. 
In figure \ref{kretsch}a we plot the  value of the Kretschman scalar invariant $K=R_{\mu\nu\alpha\beta}R^{\mu\nu\alpha\beta}$ in terms of the proper radial distance from the center of space, and observe a similar qualitative behaviour as for the flux density. The growth of the maximal value of the Kretschman scalar with the electric field is also shown in figure \ref{kretsch}b.

\section{Polarised black holes in AdS} \label{sec:BH}

In this section we consider the geometry that results from placing a black hole in the $AdS$ soliton background. The expectation is that a neutral black hole will polarize, 
pulling opposite charges to opposite hemispheres and deforming its horizon. We start with the following ansatz for a polarised black hole in $AdS$ 
\ba
 ds^{2}&=&\frac{r^2}{\left(1-r^{2}\right)^{2}}  \,   A(r,\theta)f(r)\, d\tau^{2} 
\label{BHansatz}
\\
&+&\frac{y_0^2}{\left(1-r^{2}\right)^{2}}\left[4\,\frac{G(r,\theta)}{f(r)}\, dr^{2}+C(r,\theta) \big(d\theta+2r H(r,\theta) dr\big)^{2}+B(r,\theta)\sin^{2}\theta  \, d\phi^{2} \right],\notag\\
{\cal A}&=&-ir^{2}D(r,\theta)\  d\tau\,,
\notag
\ea
where 
\be
f(r)= \left(1-r^{2}\right)^{2}-\left(1-r^{2}\right)^{3}q_0^{2}+y_{0}^{2}\left(3-3r^{2}+r^{4}\right).
\ee
The radial coordinate $r$ runs from the black hole horizon at $r=0$ to the $AdS$ boundary at $r=1$. 
Setting $A=B=C=G=1$, $H=0$ and $D=q_0$ we obtain the Reissner-Nordstrom-$AdS$ black hole of charge $q_0$, with usual radial coordinate  $y=y_0/(1-r^{2})$. We are however interested in black holes that are neutral.

Let us first pause to understand why we introduced the above constants $q_0$ and $y_0$, instead of only $y_0$. For the boundary conditions that we will impose  at $r=0$  the temperature associated with the periodicity of the thermal circle is $T=(1-q_0^{2}+3y_{0}^{2})/(4\pi y_{0})$. Setting $q_0=0$ gives the temperature of an $AdS$ Schwarzschild black hole. If we choose a seed solution with $q_0=0$ at some large temperature, then we could find solutions at lower temperatures by decreasing $y_0$ until we reached the minimum value $T_{min}$ allowed for $AdS$ Schwarzschild. However, we wish to be able to find solutions below this value, if they exist. Introducing $q_0$ allows us to do this.  First we construct a set of solutions with different values of ${\cal E}$ for some temperature above $T_{min}$. Then, to decrease the temperature and search for solutions below $T_{min}$ for non-zero ${\cal E}$, we also need to tune the parameter $q_0$. A pair $(y_0,q_0)$ is associated to some physical value of the temperature. There is here a degeneracy in the choice of such pairs, that amounts to the idea that different numerical solutions of the unknown functions at the same temperature correspond to the same geometry. We tested  this fact by comparing physical quantities like horizon area and curvature invariants of two equivalent pairs. Our choice of introducing the $q_0$ parameter was simply motivated by the Reissner-Nordstrom-$AdS$ solution for which the temperature can reach zero.

The above ansatz was chosen such that all the functions have vanishing first derivative with respect to $r$ at the horizon. In particular, we see that whatever the value of the function $A(r,\theta)$ at $r=0$, the temperature of the solution will not be affected, since it is  fixed by the parameters $y_0$  and $q_0$ in $f(r)$, as described in the previous paragraph.  As a consequence of these boundary conditions, all metric functions and the gauge field function $D(r,\theta)$ will be smooth functions of $r^2$. Given this smoothness at the horizon, the condition $A(0,\theta)=G(0,\theta)$ is fixed by the equation of motion.  This guarantees that the geometry closes smoothly at the $r=0$ axis. For the boundary conditions at the $\theta=0$ axis and the $\theta=\pi/2$ symmetric point we have chosen an ansatz such that the boundary conditions are the same as for the $AdS$ soliton of the  previous section.

Again we are left with a single boundary at $r=1$. We impose that (\ref{BHansatz}) approaches the $AdS$ boundary by setting
\be
A(1,\theta)=C(1,\theta)=B(1,\theta)=G(1,\theta)=1\,,\ \ \ H(1,\theta)=0\,.
\ee 
We also require that the gauge field approaches the dipolar potential, as for the $AdS$ soliton. This will ensure a comparison between two competing solutions with the same asymptotics. As in the previous section, we use the Einstein-deTurck trick with reference metric $A(r,\theta)=C(r,\theta)=B(r,\theta)=G(r,\theta)=1$ and $H(r,\theta)=0$ in (\ref{BHansatz}).

\subsection{Results}\label{results}

\begin{figure}[t!]
\centering
\subfloat[]{
\includegraphics[width=55mm]{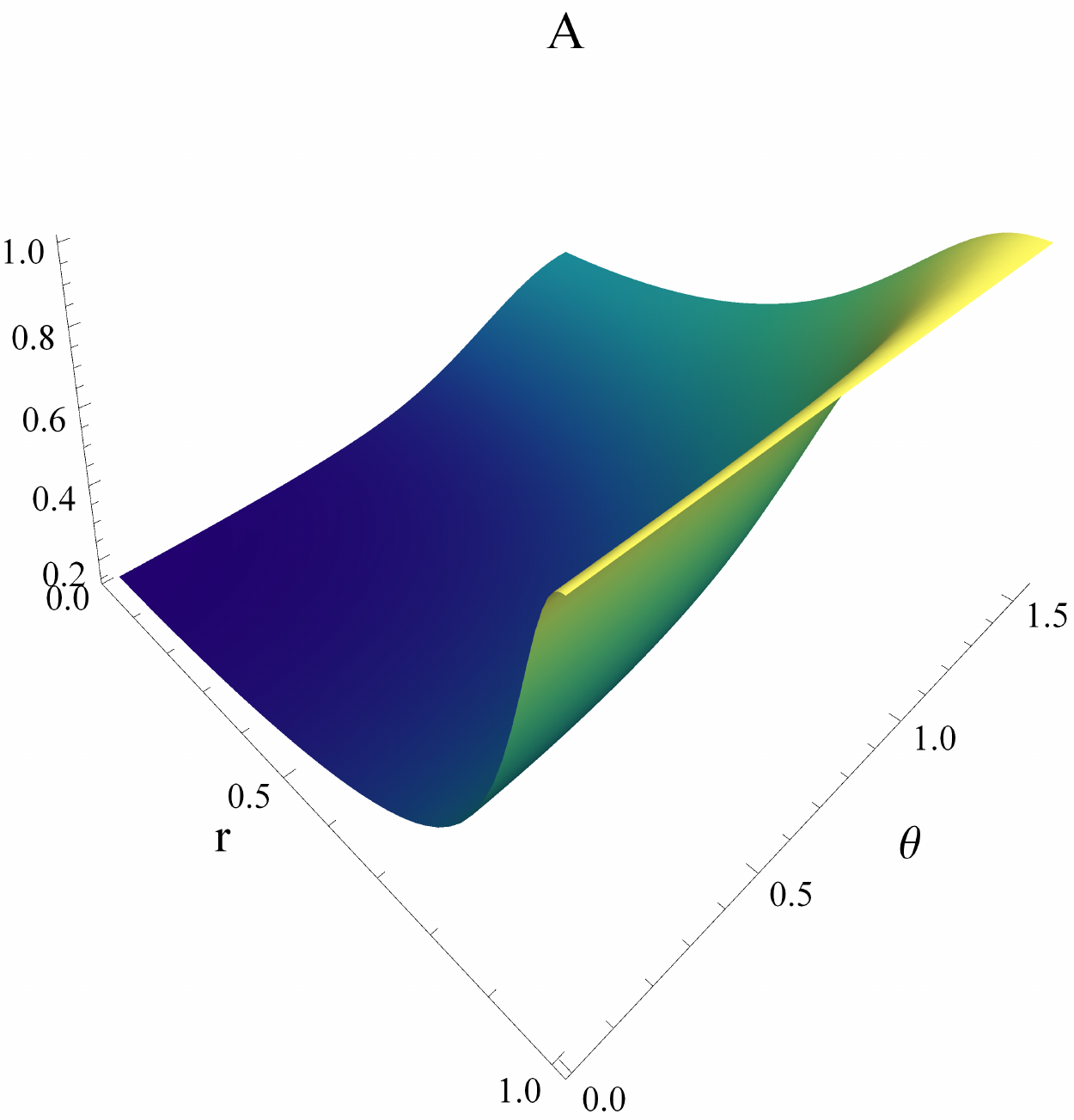}
}
\subfloat[ ]{
\includegraphics[width=55mm]{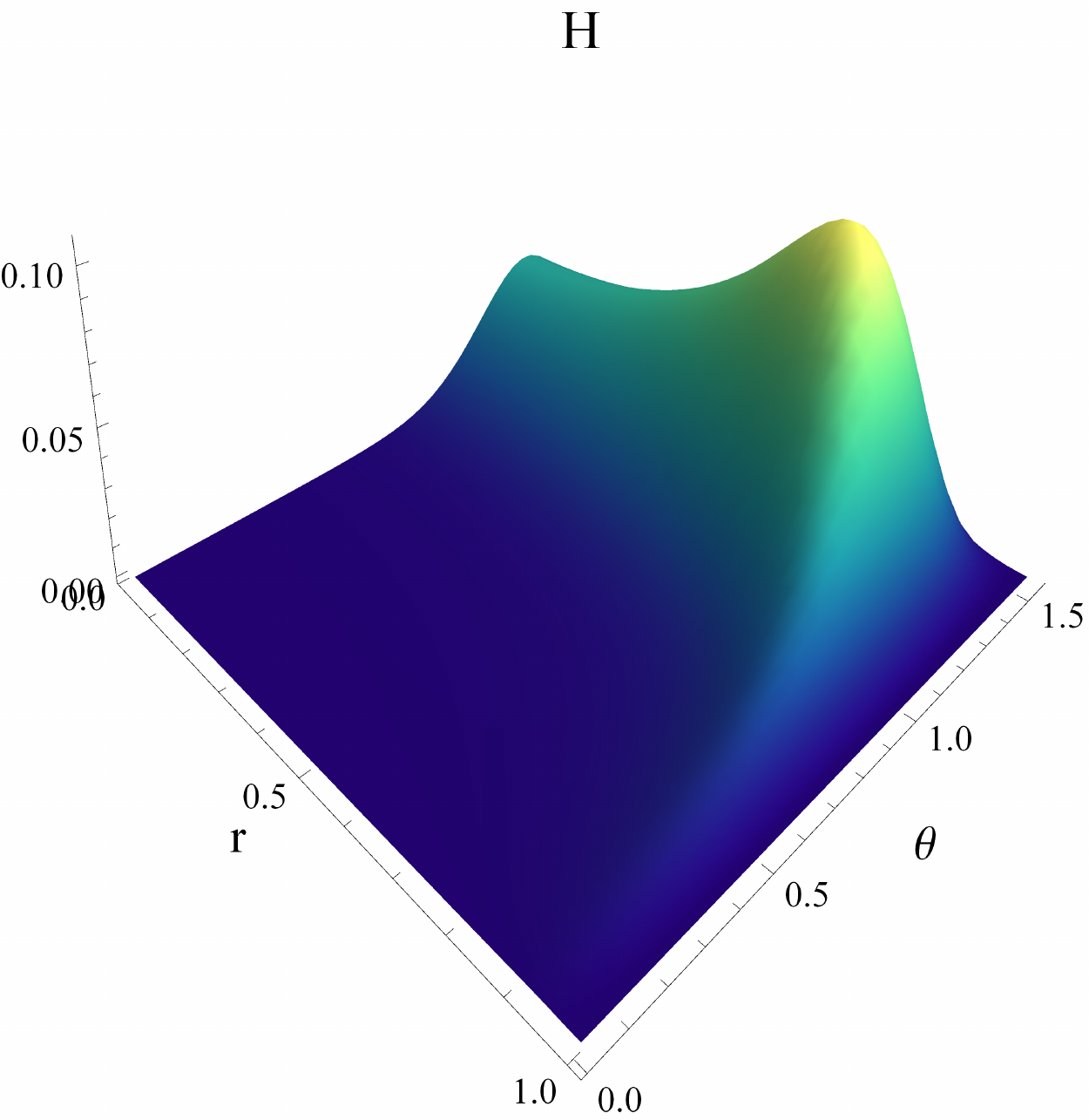}
}
\subfloat[ ]{
\includegraphics[width=55mm]{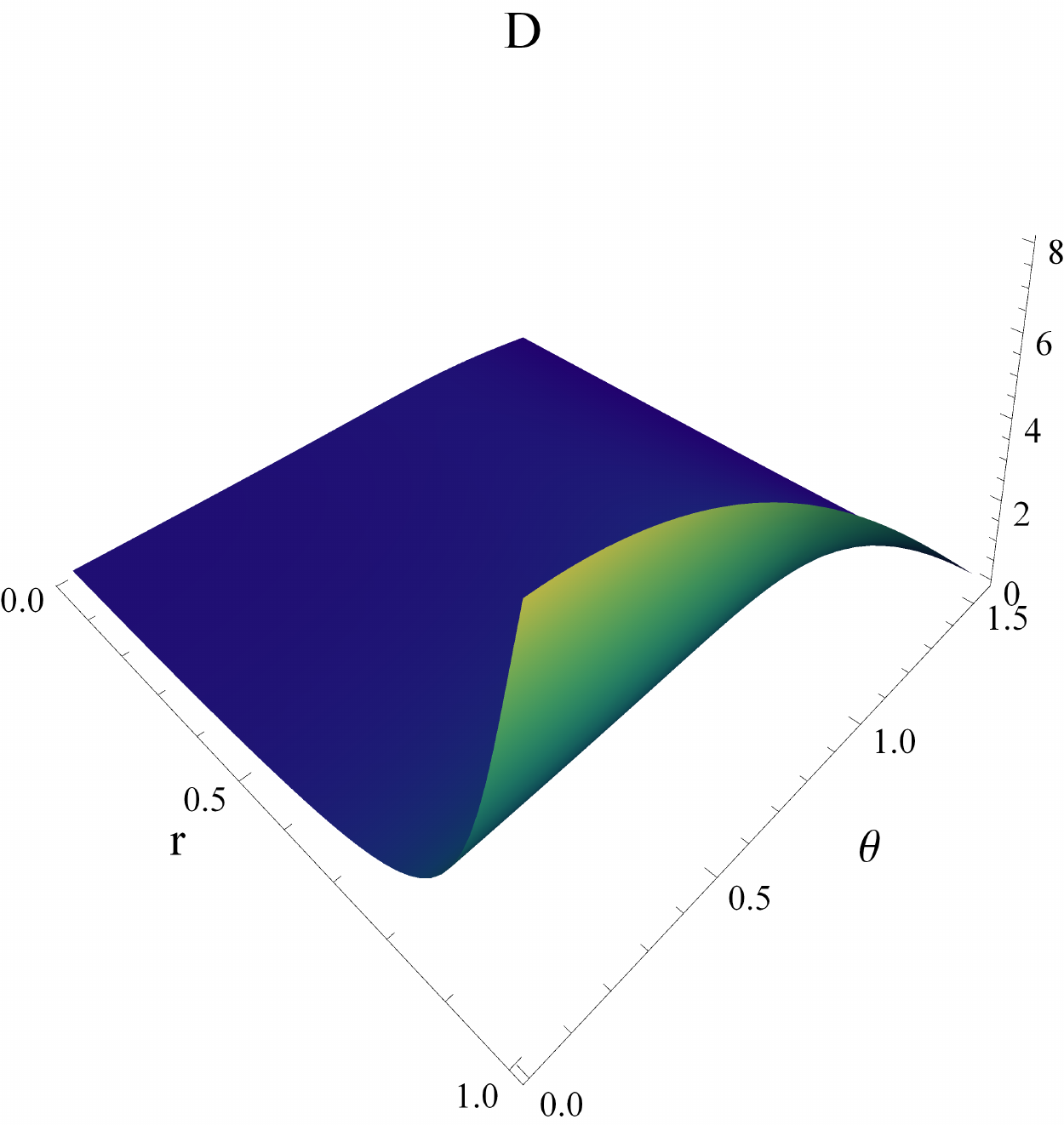}
}
\caption{Examples of numerical solutions for $A$, $H$, and $D$ for a large polarised black hole for a value of the electric field ${\cal E}=2$ and temperature $T=1/\pi$. 
In this case we used a numerical grid of $44\times 44$ points. 
\label{BH_examples}}
\end{figure}

In figure \ref{BH_examples} we show some of the functions in the ansatz (\ref{BHansatz}) for some value of the electric field and temperature. For a given temperature there will be two solutions. For that reason, we will refer to these solutions as {\em large} and {\em small} black holes. 
As we decrease the temperature in our search for solutions there is a minimum value of the temperature $T_{min}({\cal E})$ below which there is no black hole solution. This  is entirely analogous to the pure $AdS$ Schwarzschild black hole case, where now $T_{min}({\cal E})$ is a  decreasing function of  ${\cal E}$.

\begin{figure}[t!]
\centering
\subfloat[]{
\includegraphics[width=50mm]{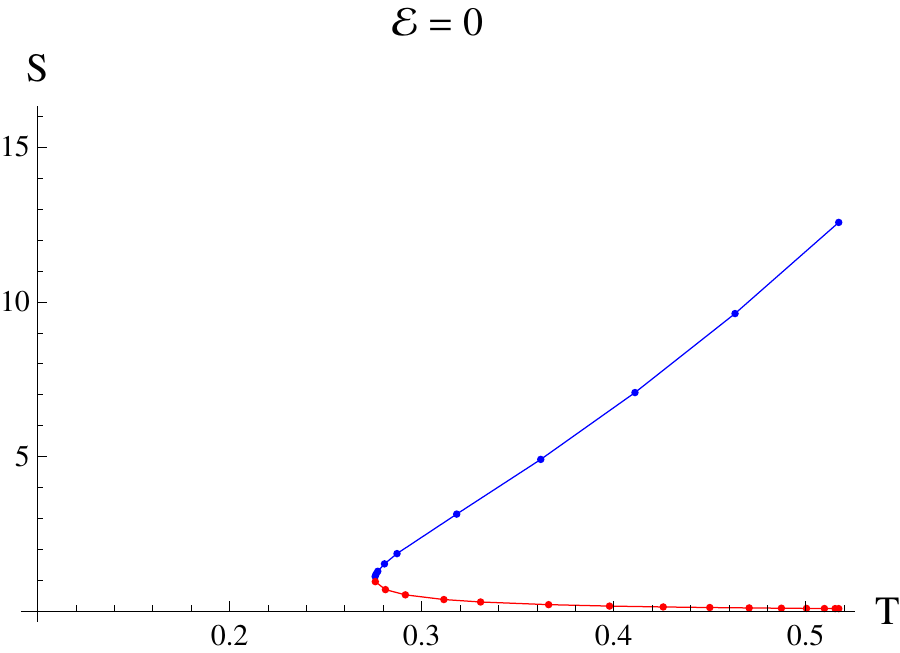}
}
\subfloat[ ]{
\includegraphics[width=50mm]{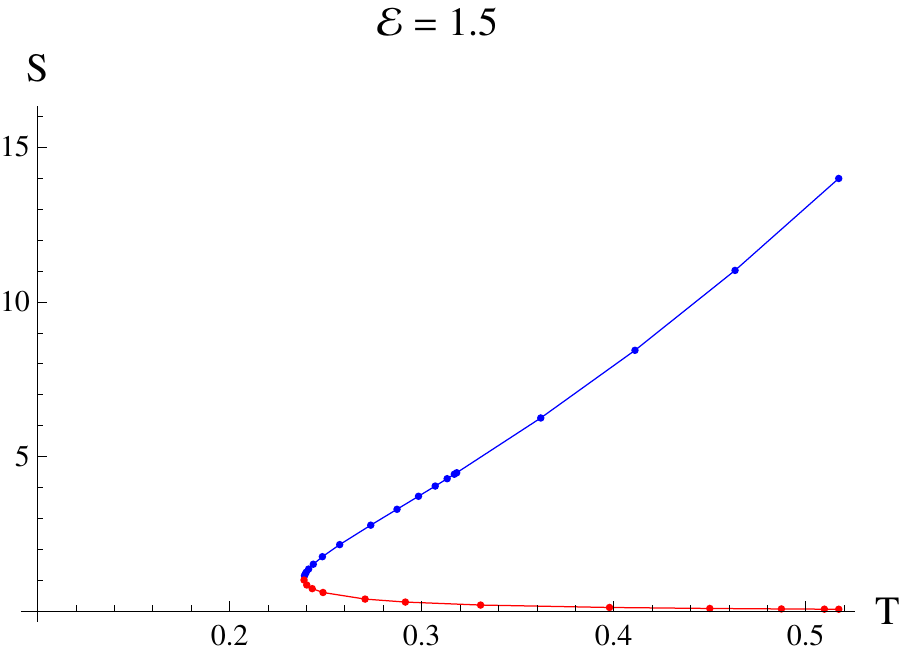}
}
\subfloat[ ]{
\includegraphics[width=50mm]{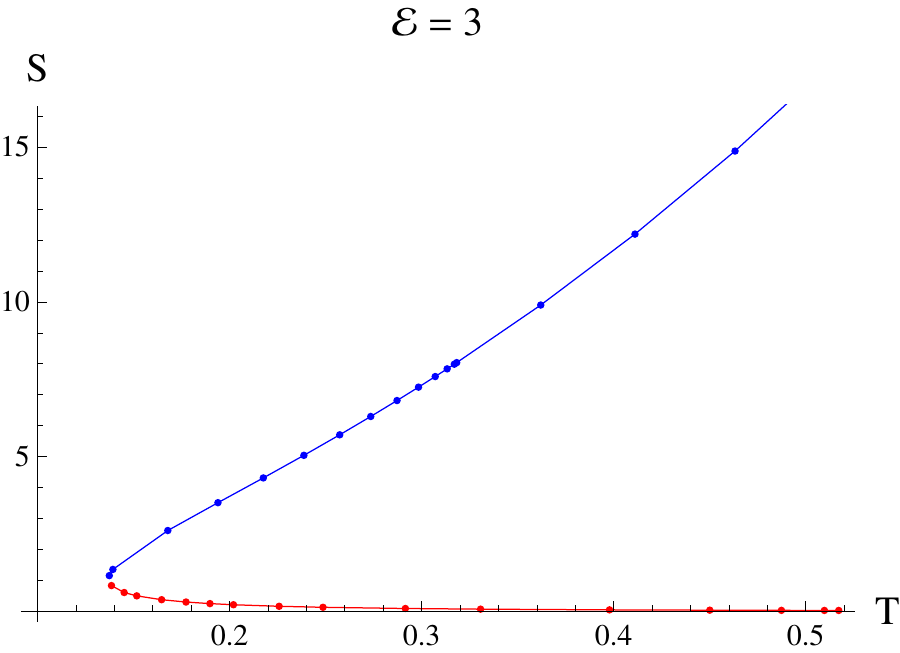}
}
\caption{Black hole entropy as function of the temperature for several values of the electric field for $G_N=1$. The blue curves correspond to the large black hole branch and the red curves to the small black hole branch. These curves meet at the minimum temperature $T_{min}({\cal E})$ below which there are no black hole solutions.
\label{entropy}}
\end{figure}

First we can look at the horizon area, which computes the entropy in the boundary theory. This is just the Hawking-Bekenstein entropy 
\ba
S=\frac{\mathcal{A}}{4 G_N}=\frac{\pi y_{0}^{2}}{G_N}\int_0^{\frac{\pi}{2}}  d\theta\, \sin\theta\sqrt{C(0,\theta)B(0,\theta)}\,.
\ea
Figure \ref{entropy} shows the  black hole entropy of the large and small black holes as a function of the temperature for several values of the electric field. The blue curves have higher entropy and therefore correspond to the large black hole branch. These large black holes grow with increasing temperature. Conversely, the small black holes, represented by the red curves, shrink with increasing temperature. We also see that the point where the curves meet corresponds to a minimum temperature  $T_{min}({\cal E})$ and that this minimum decreases with increasing ${\cal E}$.

\begin{figure}[t!]
        \centering
        \subfloat[]{
        \includegraphics[height=50mm]{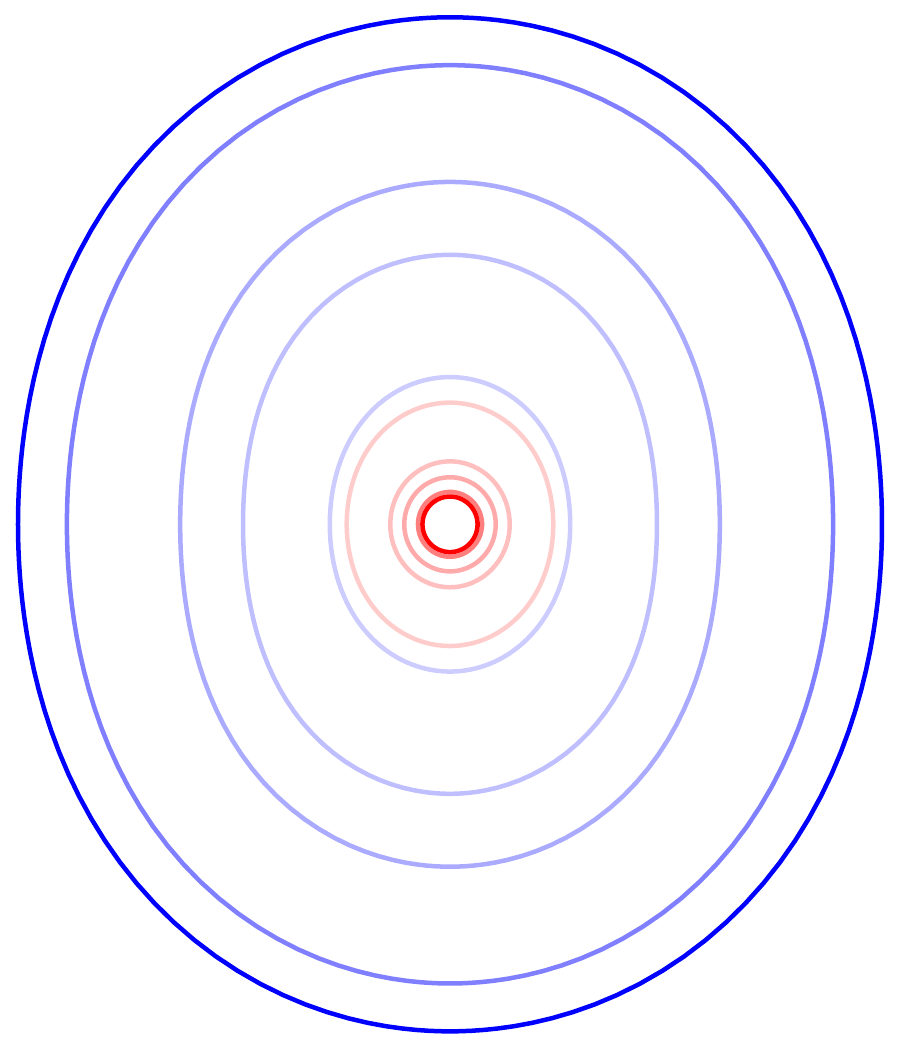}
        \label{bigvsmall}
        }
\ \ \ \ \ \ \ \ \ \ \ \ \ \ \ \ \ \ 
       \subfloat[]{
        \includegraphics[height=50mm]{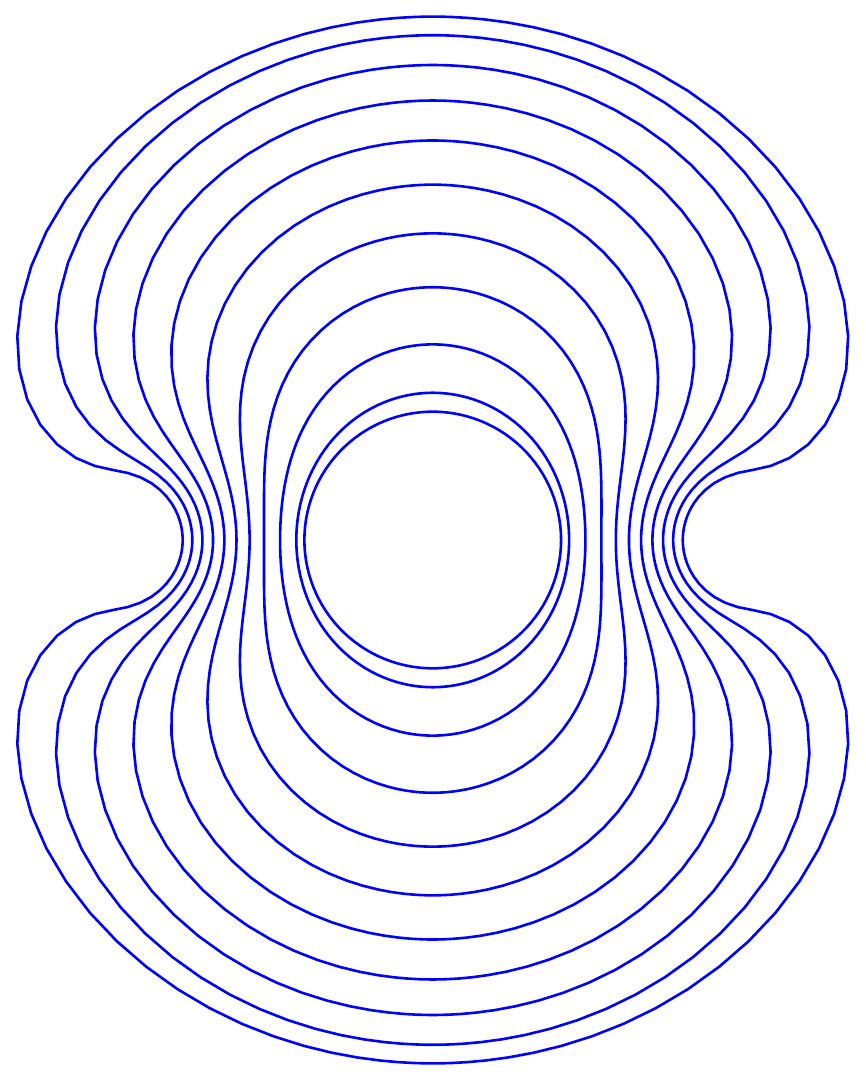}
        \label{2dplot}
        }
\caption{(a) Isometric embeddings of  black hole horizons at fixed ${\cal E}=2$. Blue curves represent  large black holes and red curves represent  small black holes. The largest blue and smallest red curves start at $T=0.52$
and get fainter as the temperature is decreased. 
The faintest blue and red lines correspond to $T=0.21$.
(b) Isometric embeddings of constant temperature black holes ($T=1/\pi$) up to a large value of ${\cal E}=10.5$.}
\end{figure}

The shape of the horizon can be drawn by computing isometric embeddings of the horizon geometry in Euclidean space. In figure \ref{bigvsmall} we plot the embeddings for the large and small black holes in blue and red, respectively. Note that the largest blue curve and the smallest red curve correspond to the same temperature. The curves get increasingly more faint as the temperature is decreased. We see that, as expected from figure \ref{entropy}, the large black hole shrinks and the small black hole expands with decreasing temperature until the minimum is reached. Figure \ref{2dplot} is a plot of the large black hole embeddings for fixed temperature up to a large value of electric field ${\cal E}=10.5$. Solutions with higher values of  ${\cal E}$ at this temperature can no longer be isometrically embedded in flat space, since its Gaussian curvature becomes too negative near the equatorial plane\footnote{This phenomenon also occurs when we look at the isometric embedding of the spatial cross section of the horizon of a rapidly rotating Kerr black hole in three dimensional flat space  - see for instance \cite{Frolov:2006yb}.}. As the electric field is increased, the black hole stretches until the geometry begins to pinch around the equator and the horizon deforms into a peanut shape. 

\begin{figure}[t!]
\centering
\subfloat[ ]{
\includegraphics[width=60mm]{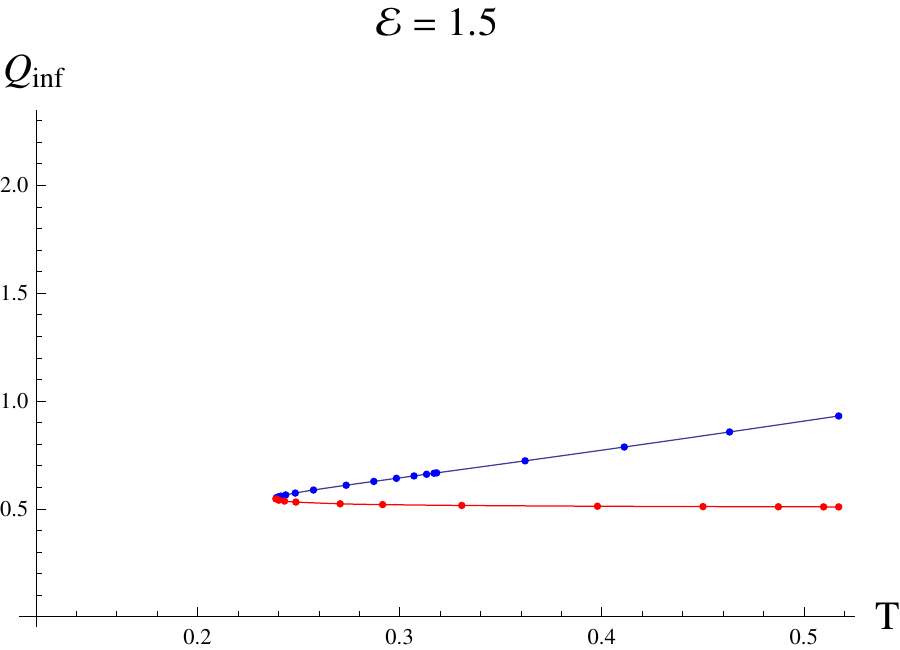}
\label{totalQa}
}
\subfloat[ ]{
\includegraphics[width=60mm]{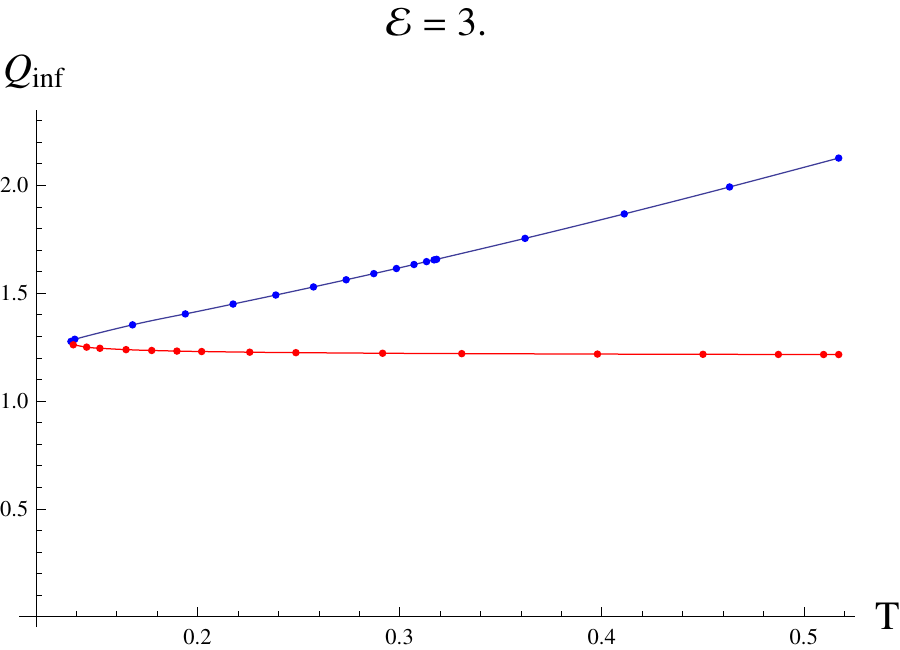}
\label{totalQb}}
\\
\centering
\subfloat[ ]{
\includegraphics[width=60mm]{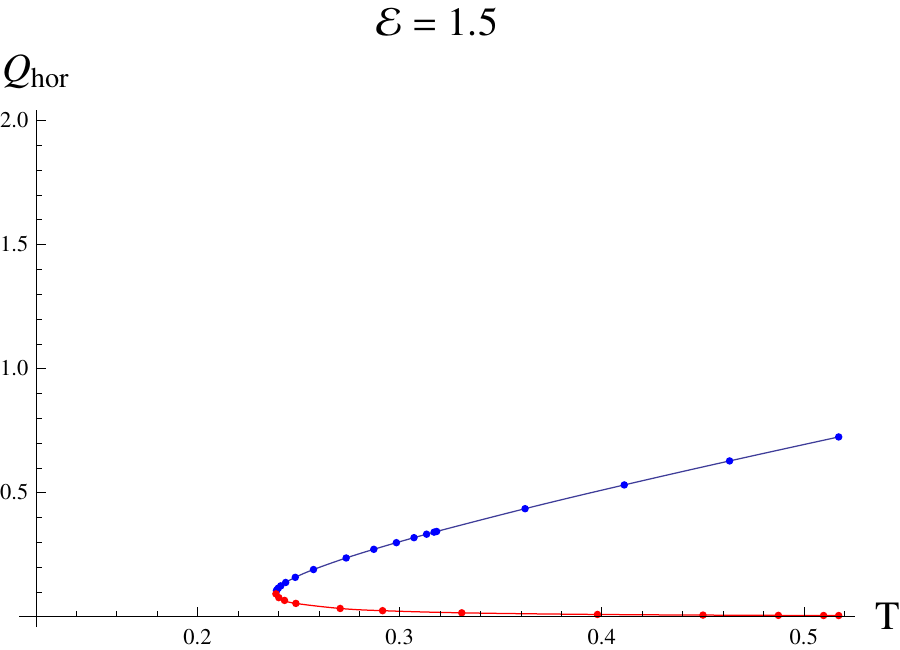}
\label{totalQc}
}
\subfloat[ ]{
\includegraphics[width=60mm]{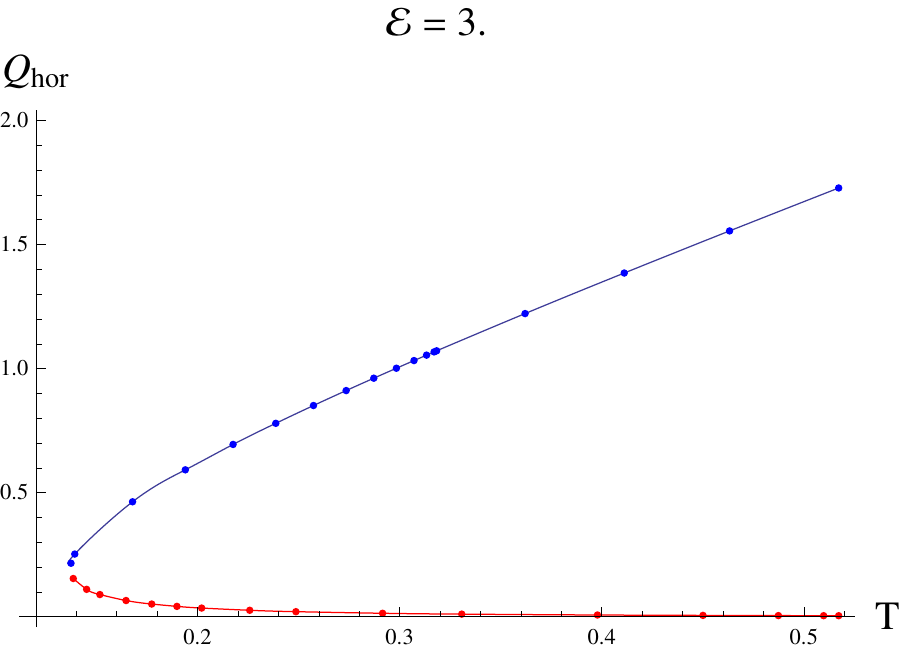}
\label{totalQd}
}
\caption{ (a)-(b) Total charge in one hemisphere  at  the $AdS$ boundary; (c)-(d) and at the black hole for $G_N=1$. The blue curves corresponds to large black holes, while the red ones to small black holes.
}
\end{figure}
Again we may study the boundary charge density $\rho(\theta)$ defined in (\ref{eq:A_asympt}) and (\ref{eq:carge_dens}). The charge is distributed along the sphere in a similar way as for the $AdS$ soliton (see figure \ref{chargedensAdS}). The total boundary charge contained in one hemisphere is shown in figures \ref{totalQa} and \ref{totalQb} for two values of ${\cal E}$. It is also interesting to look at the charge distribution and total charge that has polarised at each hemisphere of the horizon, just by computing the electric flux across the horizon. Figures \ref{totalQc} and \ref{totalQd} show the total polarised charge in each hemisphere for the same two values of ${\cal E}$. We checked that the difference between the charges in one hemisphere at the boundary and the black hole horizon equals the electric flux through the equator, as expected.

\begin{figure}[t!]
\centering
\subfloat[]{
\includegraphics[width=50mm]{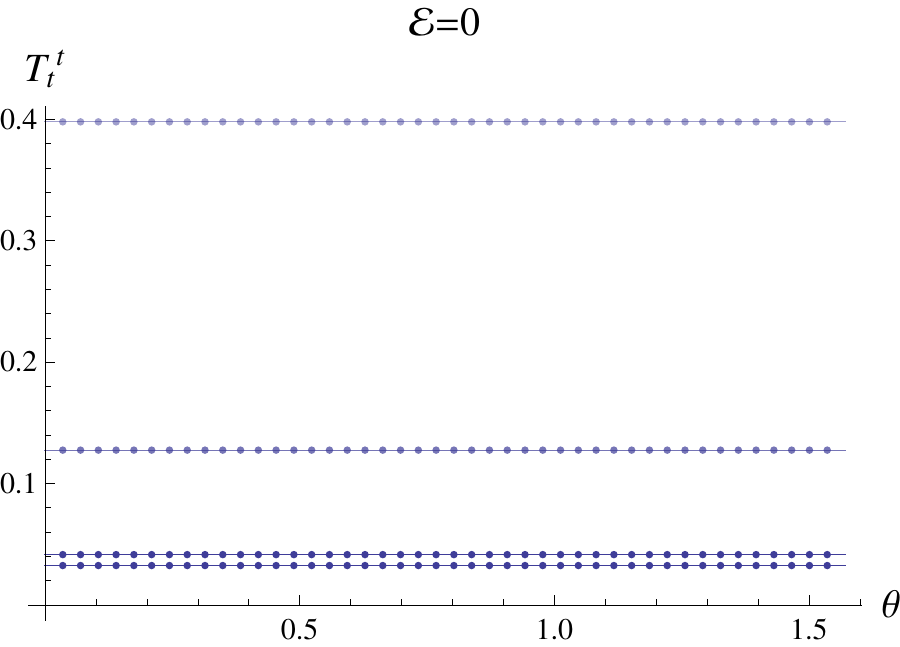}
}
\subfloat[]{
\includegraphics[width=50mm]{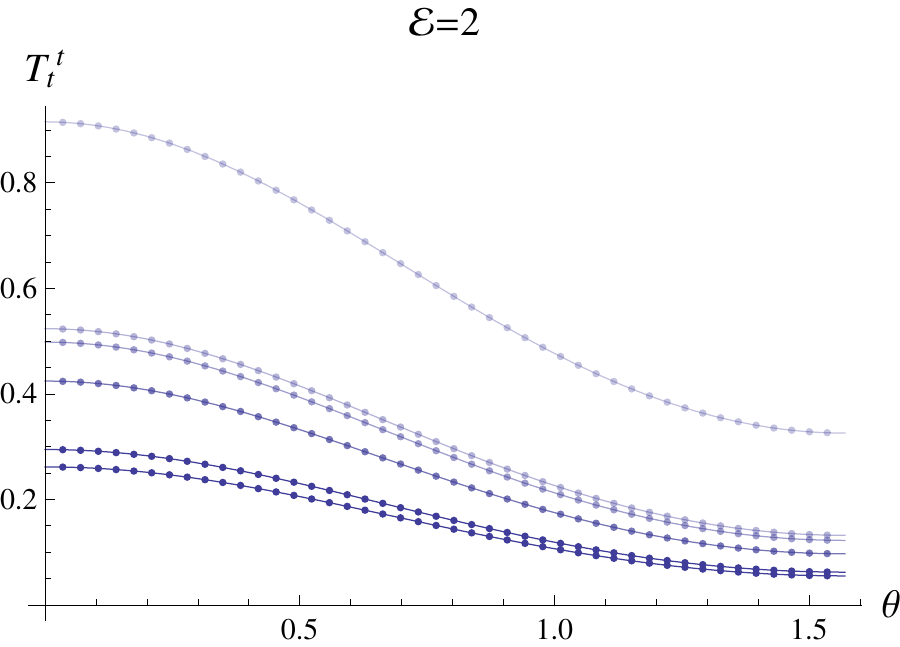}
}
\subfloat[]{
\includegraphics[width=50mm]{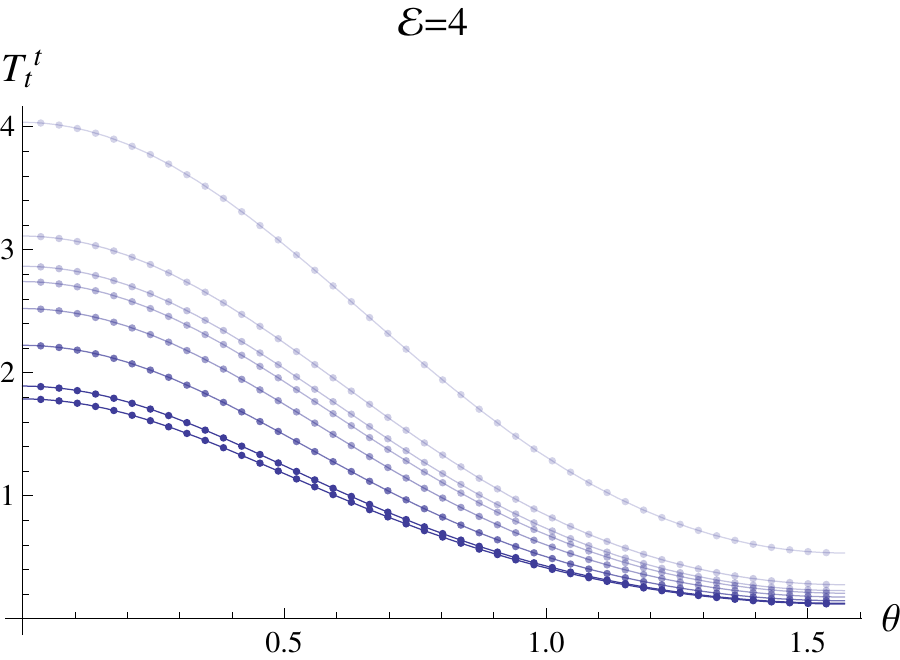}
}
\\
\subfloat[]{
\includegraphics[width=50mm]{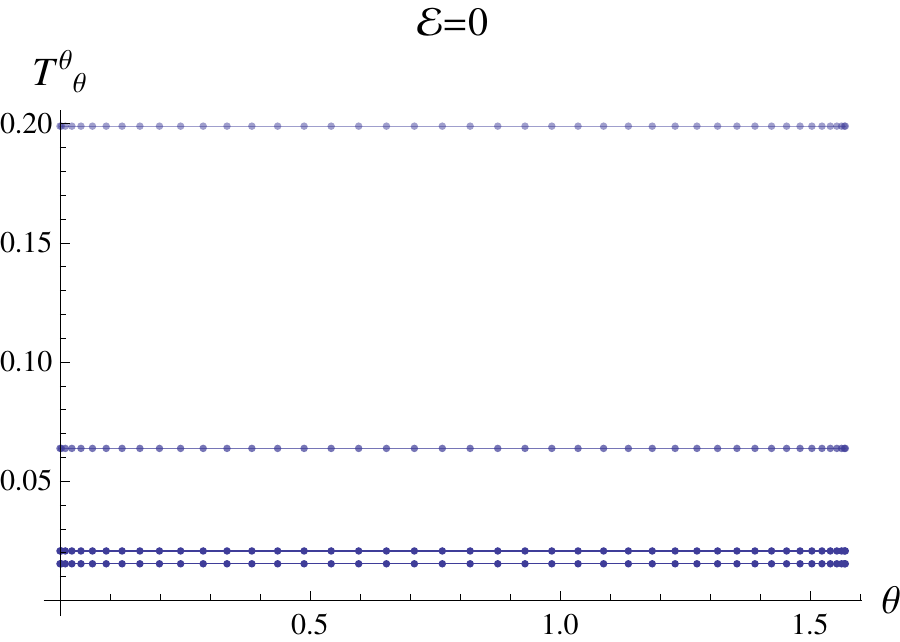}
}
\subfloat[ ]{
\includegraphics[width=50mm]{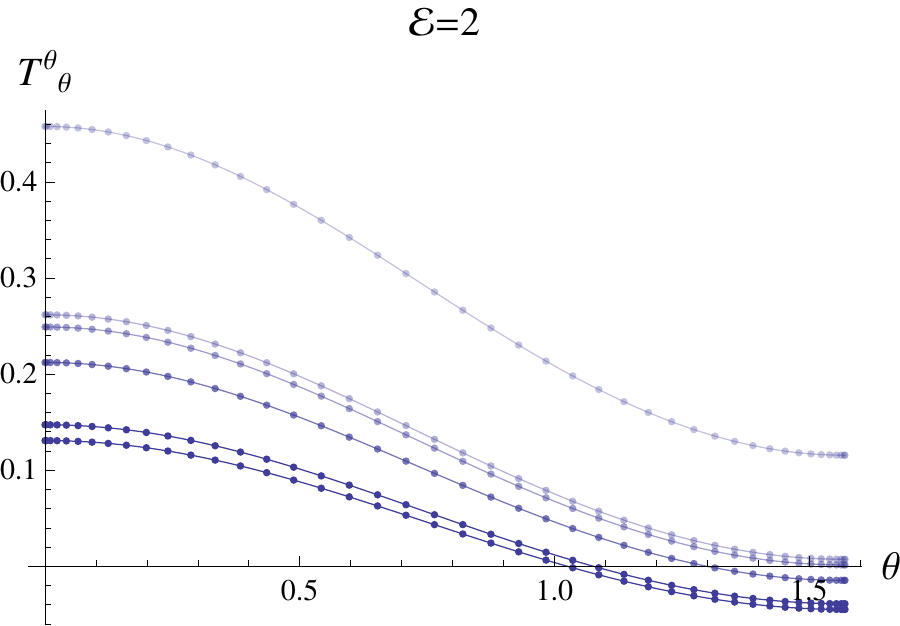}
}
\subfloat[ ]{
\includegraphics[width=50mm]{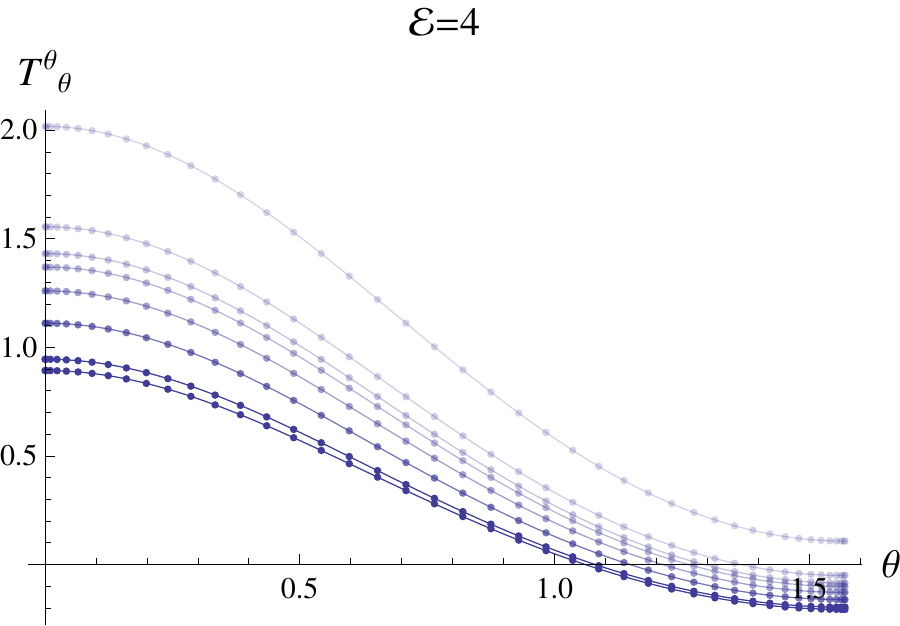}
}
\\
\subfloat[]{
\includegraphics[width=50mm]{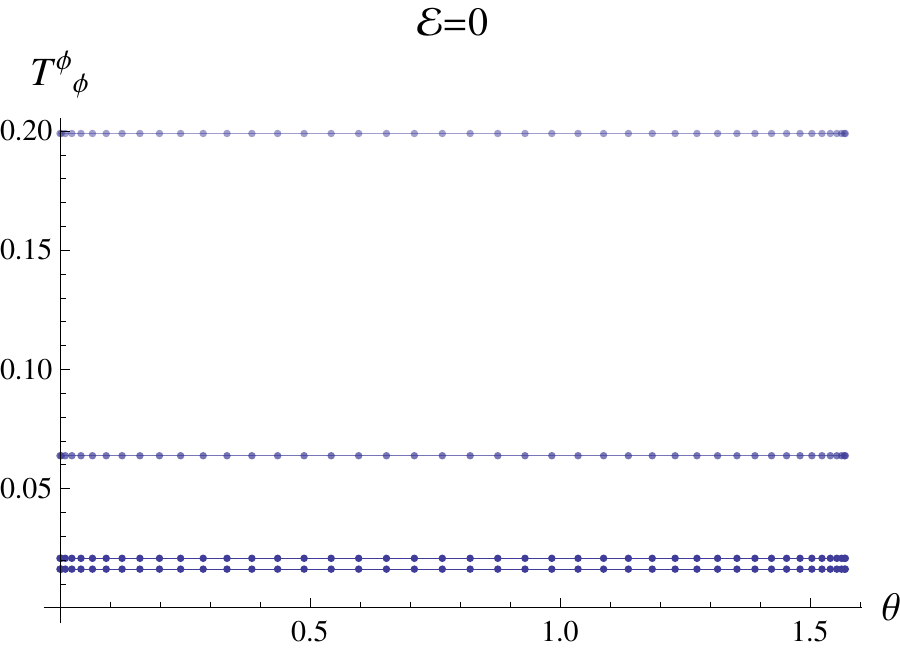}
}
\subfloat[ ]{
\includegraphics[width=50mm]{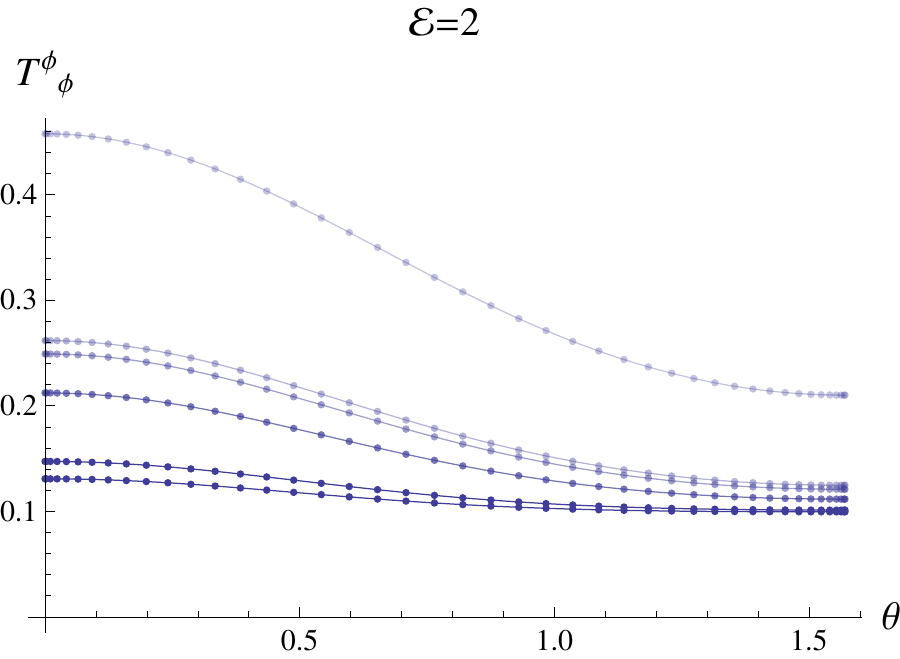}
}
\subfloat[ ]{
\includegraphics[width=50mm]{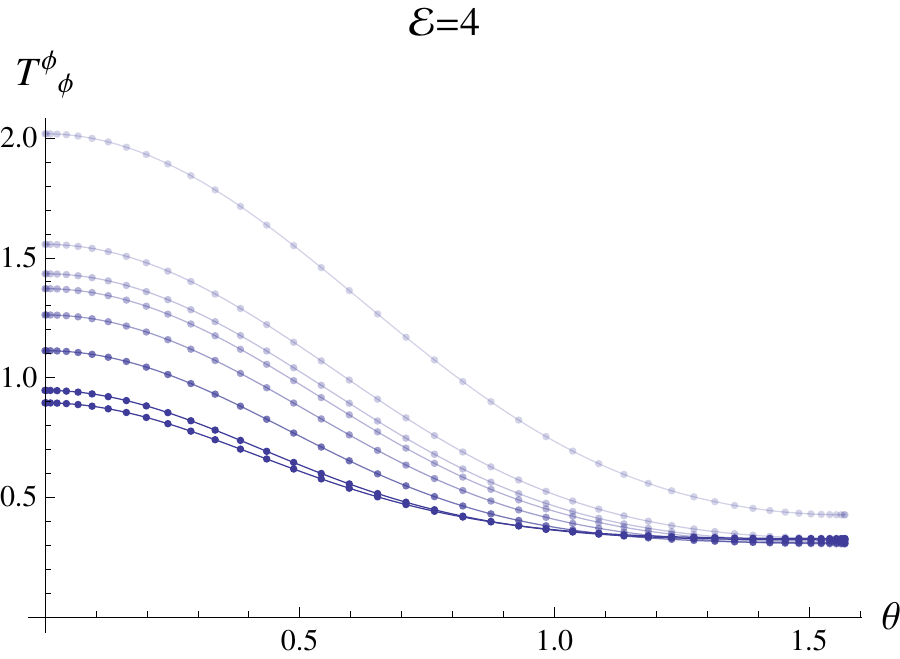}
}
\caption{ (a)-(c) Boundary energy density  for several values of ${\cal E}$ for the large
black hole; (d)-(f) the $\theta\theta$ component and (g)-(i) the $\phi\phi$  component of the boundary stress tensor at the same values of ${\cal E}$ and $T$. Darker curves correspond to higher temperatures. (Setting $G_N=1$.)
\label{StressTensorBH}}
\end{figure}

Let us now calculate the boundary energy-momentum tensor of the dual to the black hole geometry. The non-vanishing elements of the stress tensor are
\ba
T_t^{\,t}&=& \frac{y_0}{128\pi  G_N} \Big(16\left(1+q_0^2+y_0^2\right)-3y_0^2 \alpha_3(\theta)\Big),
\nonumber\\
T_{\theta}^{\,\theta}&=&\frac{y_0}{128\pi  G_N} \Big(8\left(1+q_0^2+y_0^2\right)-3y_0^2 \big(\alpha_3(\theta)+\beta_3(\theta) \big)\Big),
\\
T_{\phi}^{\,\phi}&=&\frac{y_0}{128\pi  G_N} \Big(8\left(1+q_0^2+y_0^2\right)+3y_0^2 \beta_3(\theta)\Big),
\nonumber
\ea
where, as before, $\alpha_3$, $\beta_3$, $\delta_0$ are the third and zeroth-order power-law modes associated to the functions $A$, $B$, and $D$, respectively. The profiles are similar to those found for the $AdS$ soliton, with maximum values at the pole and minima at the equator for ${\cal E}\ne0$. Some energy density profiles are shown in figures \ref{StressTensorBH}a-c for various values of the temperature and electric field magnitude. Like the $AdS$ soliton, the energy density increases with increasing ${\cal E}$. For fixed ${\cal E}$, it also increases with increasing temperature, as expected. The $\theta \theta$ component is plotted in figure \ref{StressTensorBH}d-f. It is similar to the $AdS$ soliton case, except that it now becomes negative below a critical value of $\theta$ dependent on both ${\cal E}$ and $T$. We have checked that the conservation equation (\ref{conservationT}) has a $1\%$  precision with respect to $T_\phi^{\,\phi}$ for our numerical solutions.
The $\phi \phi$ component, plotted in figure \ref{StressTensorBH}g-i, has a similar behavior. Again, this measures the pressure along $\phi$, that is independent of $\phi$, but decreases from the pole to the equator. 

\begin{figure}[h!]
\centering
\subfloat[]{
\includegraphics[width=52mm]{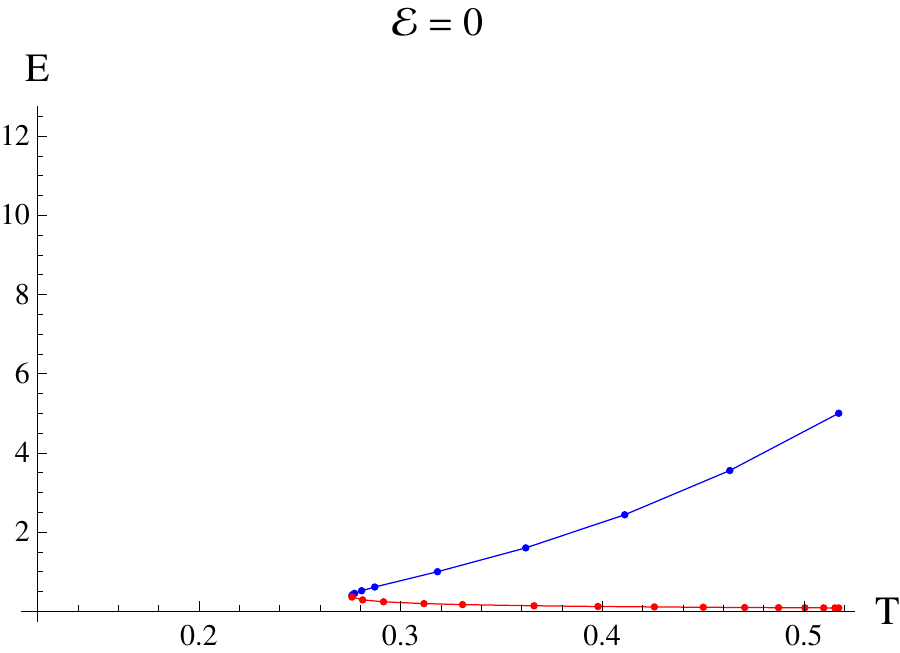}
}
\subfloat[ ]{
\includegraphics[width=52mm]{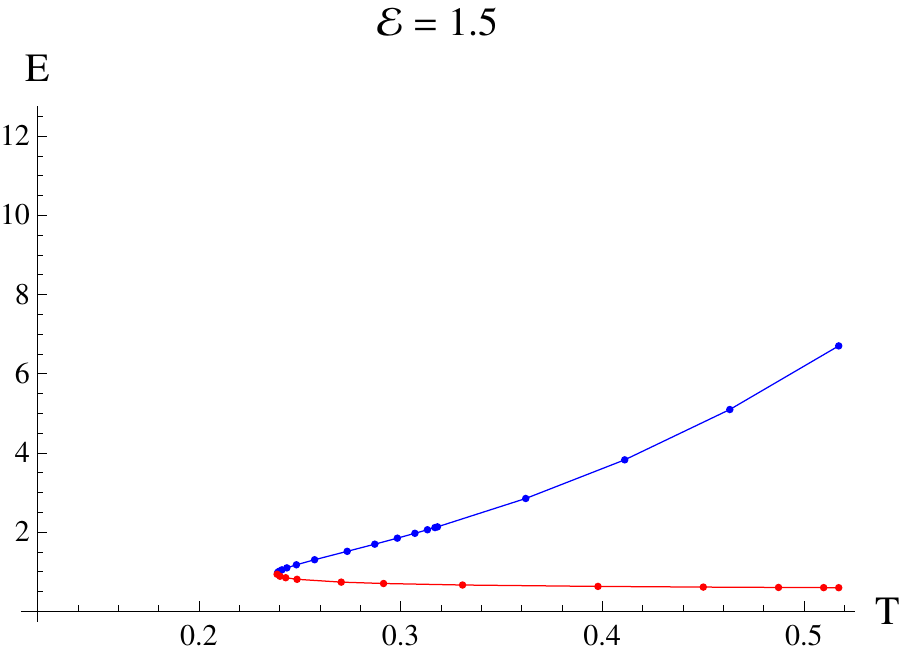}
}
\subfloat[ ]{
\includegraphics[width=52mm]{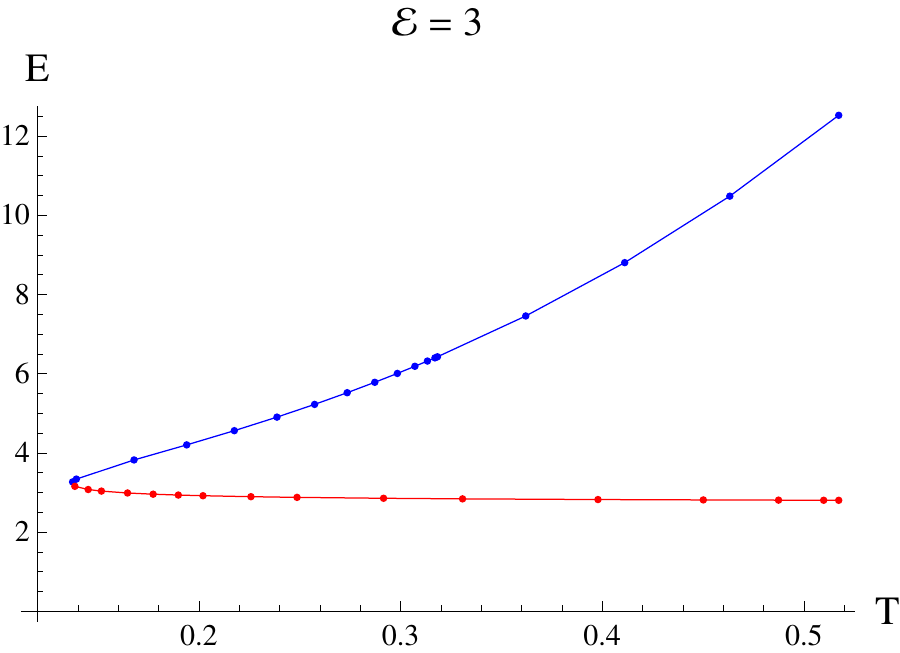}
}
\caption{Total energy at the boundary for the dual state of the black hole  for $G_N=1$.
\label{totalE}}
\end{figure}

Integrating the energy density, we get the total energy measured at infinity, which is plotted in figure \ref{totalE}. The energy of the large black hole increases with temperature while that of the small black hole decreases with temperature. Notice that both the entropy and energy agree with the analytical result of the $AdS$-Schwarzschild solution when  ${\cal E}=0$.

\subsection{Smarr formula}\label{smarr}

A nice check of the numerics is to verify whether the new black hole solutions verify the corresponding Smarr formula. This will relate boundary data with  properties of the horizon. The analysis for asymptotically $AdS$ geometries is more involved than for the asymptotically flat case because of the cosmological constant \cite{arxiv:0804.1832}.

For every killing vector $v$ we can write an antisymmetric conserved tensor
\be
(K_{v})^{ab}=\nabla^{a}v^{b}-3\omega^{ab}+2v^{[a}F^{b]c}A_{c}+2F^{ab}v^{c}A_{c}v^{a}\,,
\label{eq:FormK}
\ee
where, since $\nabla_a v^a=0$, we can define at least locally a  two-form $\omega$ such that $v^b=\nabla_a \omega^{ab}$. It is simple to see that $\omega$
can be written as 
\be
\omega^{ab}=\frac{2}{\sqrt{g}}\int\sqrt{g} \, dx^{[a}v^{b]}\,.
\ee
The $\omega$ term is necessary in $AdS$ to cancel divergences that arise from the cosmological constant term in the action. Next we need to integrate the conservation equation $d(\star K_v)=0$ from the horizon to the $AdS$ boundary to obtain
\be
\int_{hor} \star K_{v} =\int_{bound} \star K_{v}\,.
\label{eq:Smarr}
\ee
For this we need to define the form $\omega$ throughout the whole space. For the generator of time translations $v=\partial/\partial\tau$, we can choose the only non-vanishing component as 
\be
\omega^{r\tau}(r,\theta)=\frac{1}{\sqrt{g}}\left(\int_0^r\left(\sqrt{g}-\sqrt{g_{0}}\right)\,  dr'+\int_0^r\sqrt{g_{0}}\,  dr'\right).
\label{omega}
\ee
We have added and subtracted a contribution from an arbitrary metric $g_{0}$ that allows us to split up the integral into a finite and divergent piece. This is useful because it isolates the infinite piece that comes from the form $\omega$, which is canceled by the divergence that comes from the other terms in (\ref{eq:FormK}). As reference metric $g_0$ we choose the Reissner-Nordstrom-$AdS$  metric written in the coordinates of (\ref{BHansatz}) in terms of the parameters $y_0$ and $q_0$ since it has the same asymptotics as our ansatz. Making this choice, (\ref{eq:Smarr}) becomes the finite expression
\be
 TS-\frac{y_{0}^{3}}{2G_N}= E-\frac{y_{0}}{4 G_N}\left(1+y_{0}^{2}+q_0^{2}\right)-\pi \int_0^\pi  \!d\theta\sin \theta\,\rho(\theta)\Phi(\theta)-\frac{3}{4G_N}\int_0^\pi  \!d\theta \int_0^\infty \! dr (\sqrt{g} - \sqrt{g_0} )   \,.
\ee
The left-hand side is the integral taken at the horizon, while the right-hand side is taken at infinity. We computed the difference in Komar integrals at the horizon and infinity relative to the integral at infinity and found that the error is below $1\%$.

\section{Thermodynamics}\label{thermo}

Both the $AdS$ soliton presented in section \ref{sec:soliton} and the black hole of section \ref{sec:BH} were  defined with Euclidean signature. Therefore, they are already in a form that allows us to analyse the thermodynamic properties of this system. For the $AdS$ soliton we may  choose  freely the periodicity of the time circle, since the solution at fixed electric field is always the same. On the other hand, for a given temperature we need to look for the corresponding large and small black holes that are regular at the {\em Euclidean horizon}. Our goal in this section is to compare the free energies of these geometries to draw the corresponding  phase diagram.

\begin{figure}[t!]
\centering
\subfloat[]{
\includegraphics[width=50mm]{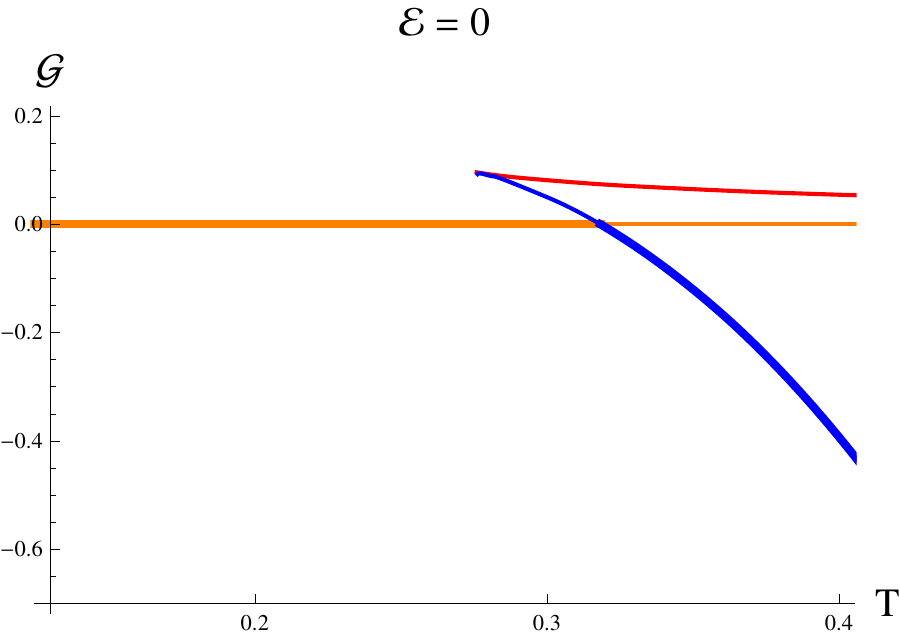}
}
\subfloat[ ]{
\includegraphics[width=50mm]{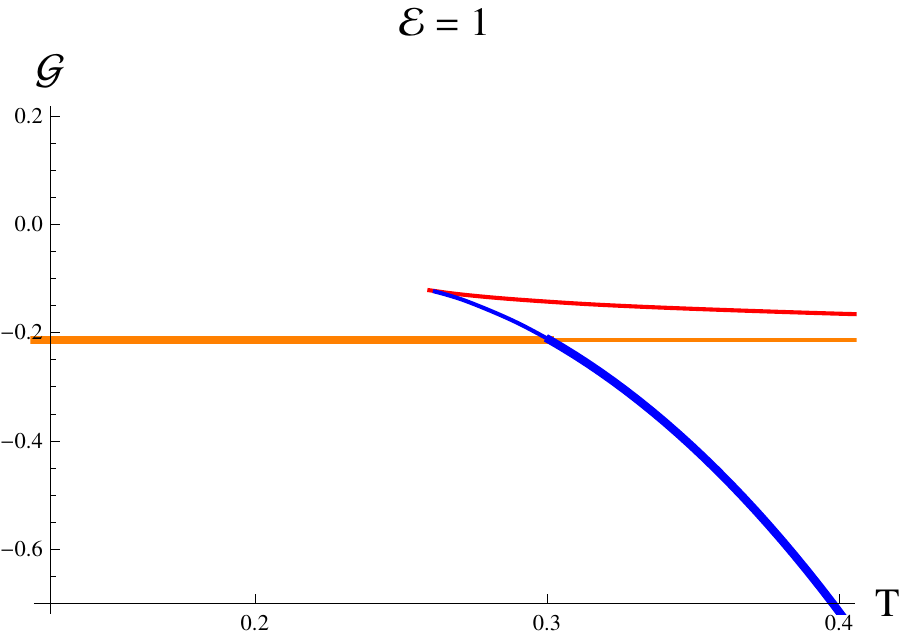}
}\\
\subfloat[ ]{
\includegraphics[width=50mm]{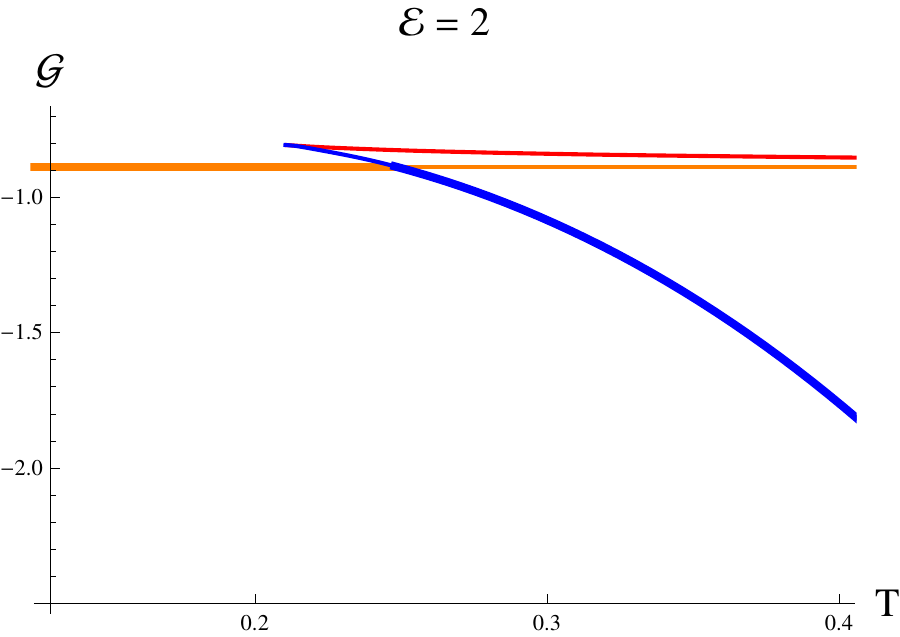}
}
\subfloat[ ]{
\includegraphics[width=50mm]{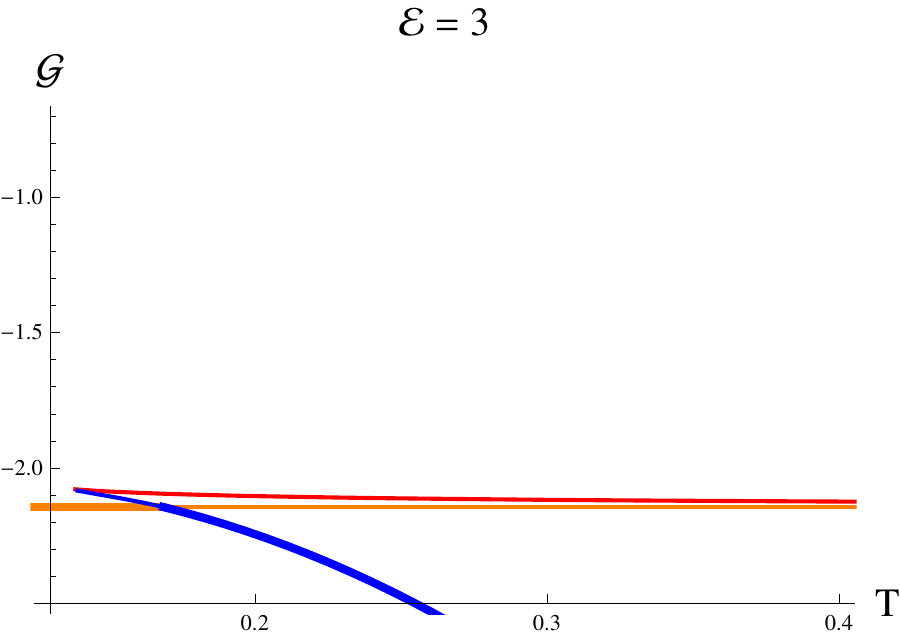}
}
\caption{Gibbs free energy for the large black hole (blue curve), small black hole (red curve) and $AdS$ soliton (orange line) for several values of the electric field as a function of the temperature. Thicker lines single out the dominant phase. In these plots we set $G_N=1$.
\label{free2d}}
\end{figure}

\begin{figure}[t!]
\centering
\includegraphics[width=80mm]{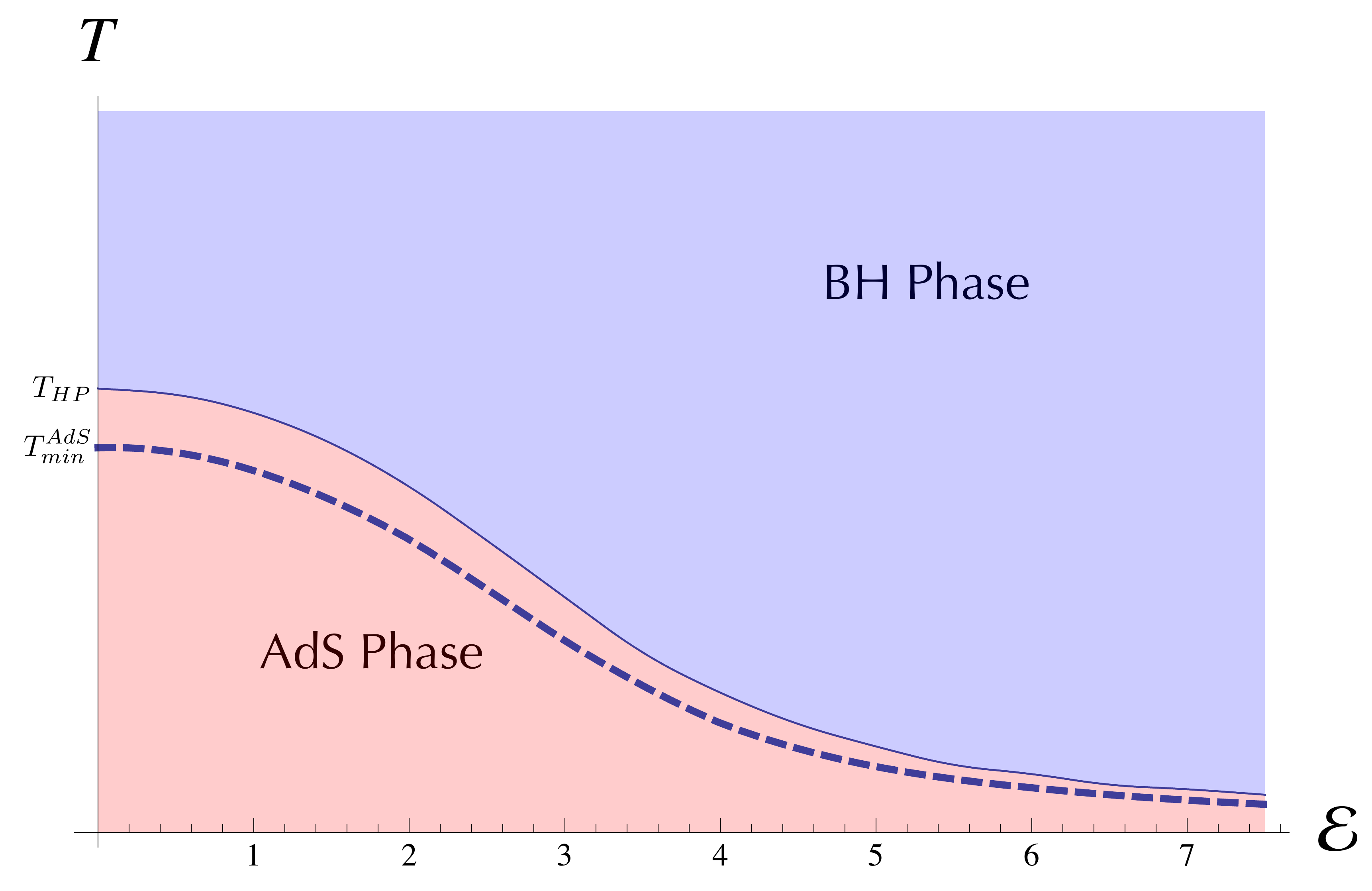}
\caption{Phase diagram with the critical temperature (solid curve) above which the black hole phase is thermodynamically favoured. The minimum  temperature for the black holes  (dashed curve) is also shown. 
 \label{phasediagram}}
\end{figure}

The free energy associated to a geometry with an electrostatic source $\Phi(\theta)$ at the boundary is 
\be
{\cal G}= E-TS-2 \pi \int_0^\pi  d\theta \sin \theta\,\rho(\theta) \, \Phi(\theta) \,,
\ee
where $E$ is the energy, $S$ is the Bekenstein-Hawking entropy and $\rho(\theta)$ is the charge density at the boundary. For the $AdS$ soliton the entropy vanishes since there is no horizon, and the energy and charge density  are independent of $T$.  For both the $AdS$ soliton and black hole geometries all functions were determined in the previous sections. Figure \ref{free2d} shows the Gibbs free energy of the large black hole (blue curve), the small black hole (red curve), and $AdS$ soliton (orange line). The curves for the $AdS$ soliton and black hole phases are thicker when they are the dominant phase. The large and small black hole picture is very similar to that of the RN-$AdS$ black holes in a constant potential background  in \cite{hep-th/9902170,hep-th/9904197}. The small black hole branch always has higher free energy than the large black hole branch and is therefore not thermodynamically favoured. The phase transition between the large black hole and $AdS$ soliton phases occurs when the curves of the two phases cross. We see that for ${\cal E}=0$ the phase transition occurs at $T=1/\pi$, in agreement with the first order Hawking-Page phase transition of a Schwarzschild black hole in $AdS$. As the electric field increases, the phase transition moves toward lower temperatures.  Figure \ref{phasediagram} shows the phase diagram. The blue region is in the black hole phase while the red region is in the $AdS$ soliton phase. It seems that the critical temperature will tend towards zero as the electric field becomes very large. The lower dashed curve marks the minimum temperature of the black hole solutions, where large and small black holes meet. 

To analyze the stability of the black holes, we calculated the specific heat $C=T (dS/dT)$. We find that large black holes are stable, while small black holes are not, as follows from figure \ref{entropy}. It is also clear from figure  \ref{entropy} that the specific heat will diverge when $T\to T_{min}$.

\subsection{Free boson on a sphere  with dipolar potential}
 
One very simple toy model to gain intuition about the dual field theory is  
a free charged scalar field satisfying the Klein-Gordon equation on the two-sphere
\be
D_{\mu}D^{\mu}\varphi=M^{2}\varphi\,,
\ee
where $D_{\mu}=\nabla_{\mu}-iC_{\mu}$. 
We shall focus on the conformal case that corresponds
to $M^{2}=1/4$ with units such that the sphere radius is equal to 1. Considering the geometry of the cylinder $S^{2}\times\mathbb{R}_{t}$
and the dipolar potential $C={\cal E}\cos\theta \,dt$, we find 
\be
\left[-\left(\partial_{t}-i{\cal E}\cos\theta\right)^{2}+\nabla_{S^{2}}^{2}\right]\varphi=M^{2}\varphi\,.
\ee
For ${\cal E}=0$ the eigenstates  are simply given by the spherical harmonics $|l,m\rangle$ with wave-functions
\be
\langle \theta,\phi|l,m\rangle\propto e^{im\phi}P_l^m(\cos \theta)\,,
\ee
where $P_l^m$ are the associated Legendre polynomials. The  spectrum of the Hamiltonian $\hat{H}=i\partial_t$ is given by
\be
\hat{H} \,|l,m\rangle= \sqrt{l(l+1)+M^2 }\,|l,m\rangle =\left(l+\frac{1}{2} \right)|l,m\rangle\,,
\ee
in agreement with the scaling dimensions of the local operators $\partial_{\mu_1}\dots \partial_{\mu_l}\varphi$ that constitute the conformal family of a free scalar field $\varphi$ in three dimensions. These states have charge 1 and are degenerate with the states of charge $-1$ associated with $\partial_{\mu_1}\dots \partial_{\mu_l}\bar{\varphi}$.

\begin{figure}[t!]
\centering
\subfloat[]{
\includegraphics[width=65mm]{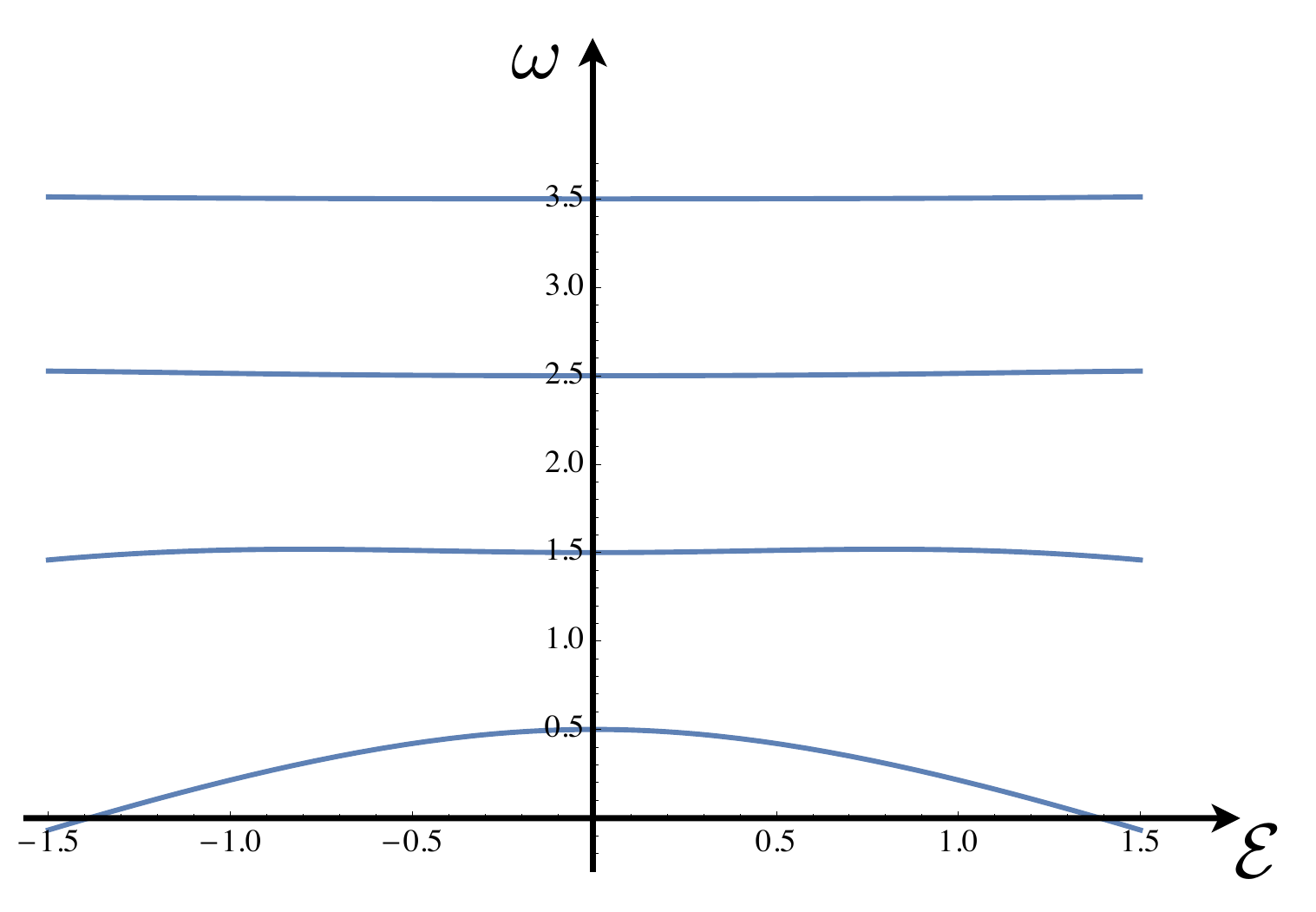}
}
\subfloat[ ]{
\includegraphics[width=65mm]{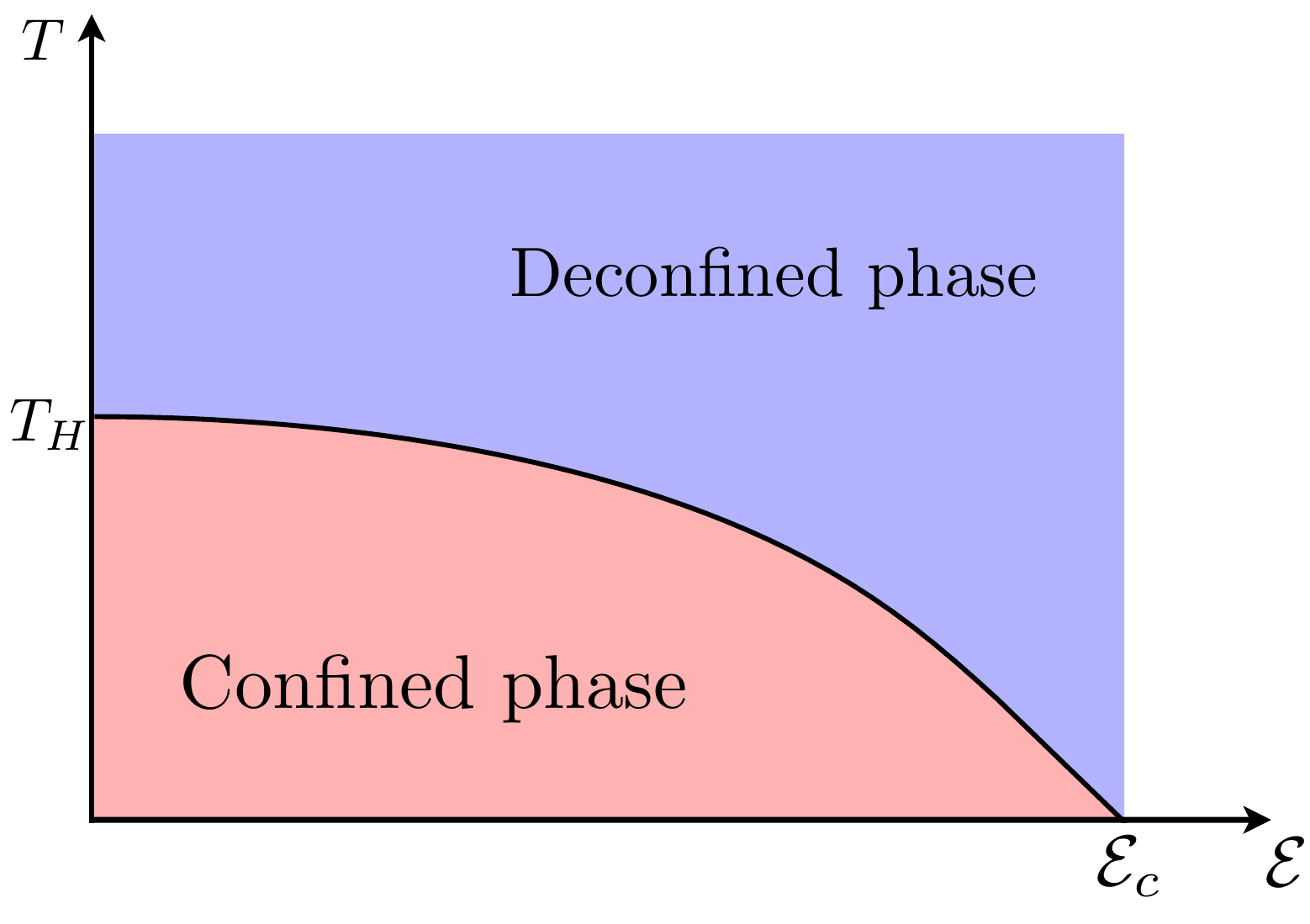}
}
\caption{(a) First four energy levels in the $m=0$ sector as a function of the dipolar potential $\cal E$. (b) Large $N$ phase diagram of a free  adjoint scalar field. The Hagedorn temperature decreases with   the dipolar potential ${\cal E}$ and goes to zero as ${\cal E}\to{\cal E}_c$.
 \label{fig:Hagedorn}}
\end{figure}

After turning on the electric field ${\cal E}$, the hamiltonian becomes
\be
H=\sqrt{M^2-\nabla^2_{S^2} }-{\cal E} \cos \theta\,,
\ee
which is still diagonal in the azimuthal quantum number $m$ but it becomes an infinite tridiagonal matrix in the quantum number $l\ge |m|$,
\be
\langle l',m'|H|l,m\rangle=\delta_{m,m'}\left\{ \delta_{l,l'}\left(l+\frac{1}{2} \right)+{\cal E}\left[\delta_{l,l'-1}\,\sqrt{\frac{(l+1-m)(l+1+m)}{(2l+1)(2l+3)}}+(l\leftrightarrow l')\right]\right\} .
\label{Hamiltonian}
\ee
For each $m$ one can truncate the values of $l=|m|,|m|+1,\dots$ up to a maximal value $L$ and (numerically) compute the eigenvalues of the resulting finite matrix.\footnote{The lowest eigenvalues of the truncated matrices converge exponentially fast to the eigenvalues of the infinite matrix as we increase the cutoff $L$.}  The low energy spectrum in the $m=0$ sector is shown in figure \ref{fig:Hagedorn}a. Notice that for ${\cal E}>{\cal E}_c\approx 1.3868$ the single particle ground state energy becomes negative. The energy decreases because the wave-function concentrates around the   pole   where the potential is negative. In fact, the single-particle states always come in degenerate pairs labelled by the charge 1 or $-1$, with wave-functions related by the interchange  $\theta \leftrightarrow \pi -\theta$. For ${\cal E}>{\cal E}_c$ the system is unstable because we can lower the energy without bound by accumulating bosons in the single-particle ground states.

Let us now consider the system at finite temperature. We can define the single-particle partition function
\be
z(x,{\cal E})= 2\sum_{m=-\infty}^\infty\sum_{k=0}^\infty e^{-\beta \omega_{m,k}({\cal E})}\,, 
\qquad \qquad x\equiv e^{-\beta}\,,
\ee
where $\omega_{m,k}({\cal E})$ is the energy spectrum of the hamiltonian (\ref{Hamiltonian}) and the factor of 2 accounts for the two possible charge assignments. We would like to consider a $SU(N)$ gauge theory with a scalar field $\varphi$ in the adjoint representation. Then, the gauge-invariant states can be written as products of traces of products of the elementary fields $\varphi$ and $\bar{\varphi}$. As explained in \cite{hep-th/9908001, hep-th/0110196, hep-th/0310285}, in the large $N$ limit and below the Hagedorn temperature, the  full partition function is given by
\be
\log Z(x ) = -\sum_{n=1}\log \big[
1-z_B\!\left(x^n \right)+(-1)^n z_F\!\left(x^n \right)
\big]\, ,
\ee
where $z_B$ and $z_F$ are the bosonic and fermionic single-particle partition functions. For simplicity, we will only consider the bosonic contribution $z(x,{\cal E})$ of the complex scalar field $\varphi$. The Hagedorn temperature is then given by the condition 
\be
z(x_H,{\cal E})=1\,,\qquad\qquad x_H=e^{-\beta_H}\,.
\ee
We plot the corresponding phase diagram in figure \ref{fig:Hagedorn}b. As usual, there is a low temperature confined phase and a high temperature deconfined phase separated by a Hagedorn phase transition. The novelty is that the Hagedorn temperature decreases with the dipolar potential ${\cal E}$ and goes to zero as ${\cal E}\to{\cal E}_c$.

\section{Conclusion}

\begin{figure}[t!]
\centering
\subfloat[]{
\includegraphics[height=65mm]{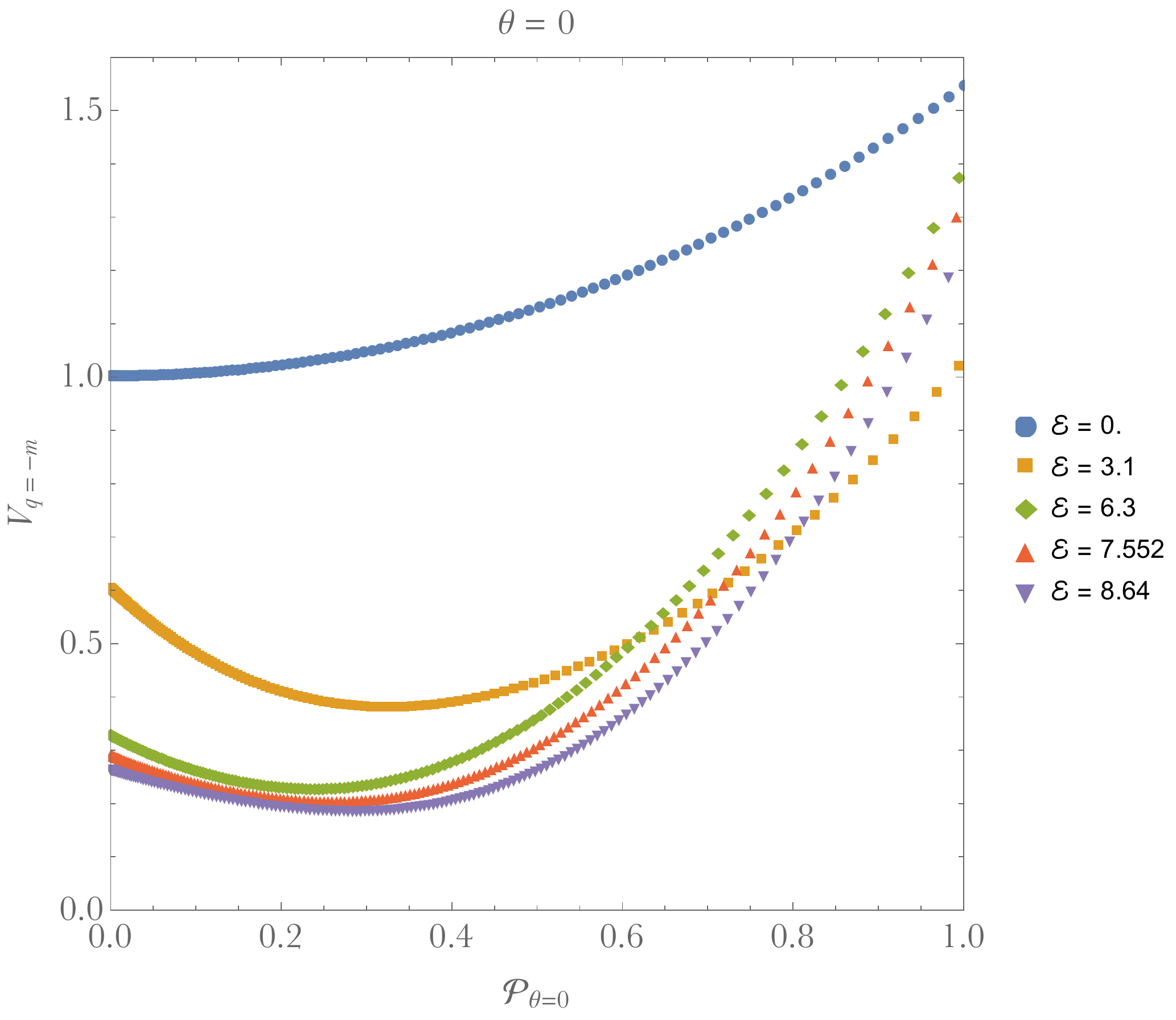}
}
\subfloat[]{
\includegraphics[height=65mm]{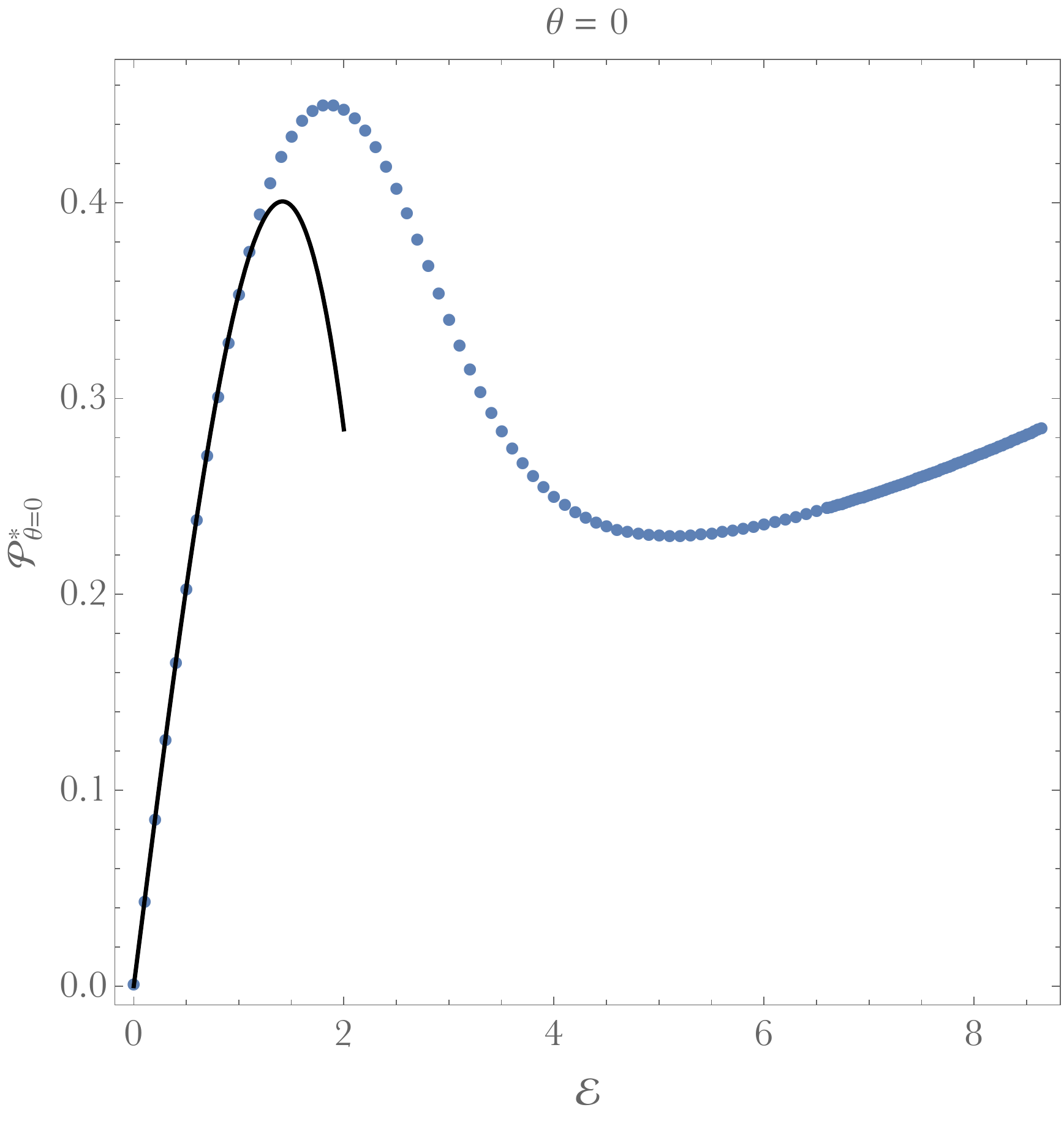}
}
\caption{Potential for timelike static orbits of extremal charged particles in the $AdS$ background is plotted in (a) for several values of the electric field. In (b) we plot the location of the extremal charged particles as a function of the electric field. The black curve is the result obtained via perturbation theory analysis, which is detailed in Appendix \ref{ap:I}.
\label{geo}}
\end{figure}

Given the shape of the horizon of the polarised black hole for large dipolar potential, it is natural to wonder if the thermodynamically favourable solution contains two black holes. 
In order to investigate this question we consider the free energy variation of the system when adding two infinitesimal small black holes of opposite charges.
The first step is to determine their equilibrium position. Since the putative black holes are infinitesimal we can work in the probe approximation.
Therefore, equilibrium positions are just located at  the minima of the potential
\be
V=\sqrt{g_{\tau\tau}}-\frac{q}{m}A_\tau\,,
\ee
where $q$ and $m$ are the charge and mass of the black holes.
Since these black holes are small, their properties can be described by Reissner-Nordstrom black holes in flat space. These have
\be
m=r_+-r_+^2 2\pi T\,,\qquad
q^2=r_+^2-r_+^3 4\pi T\,,\qquad
s=\pi r_+^2\,,
\ee
where $r_+$ is the outer horizon radius and $s$ is the entropy (we are using $G_N=1$).
At finite temperature $T$, we find $q/m \to 1$ in the probe limit $r_+\to 0$.
We find that timelike static orbits of oppositely charged test particles do exist  along the $\theta=0$ axis for $q=m$ in  the $AdS$ soliton background.
The potential and position of the corresponding  minima  are plotted in figure \ref{geo}. 
We may now consider the   free energy variation of the solution corresponding to the addition of a probe extremal black hole,
\be
\delta {\cal G} = m\left(\sqrt{g_{\tau\tau}}-\frac{q}{m}A_\tau\right)_{min}   -T s =  r_+ 
V_{min}+ O(r_+^2)  \,.
\ee
We conclude that it is not advantageous to add probe black holes if the probe potential at the minimum is positive, which is what we found at least up to ${\cal E} \sim 9$.
It would be interesting to look for new two black hole solutions, beyond the probe approximation, to complete the phase diagram. This could lead to a maximum value of the
electric field above which the $AdS$ soliton is not the favoured low temperature phase. 

There are some similarities between the weak coupling phase diagram of figure \ref{fig:Hagedorn}b and the gravitational (strong coupling) phase diagram of figure \ref{phasediagram}. It would be interesting to study in detail these phase diagrams in a concrete realisation of gauge/gravity duality like ABJM theory \cite{arXiv:0806.1218}. In that case, one can interpolate between the two phase diagrams by changing the t'Hooft coupling of the theory. The main qualitative difference between  weak and strong coupling seems to be the existence, or not, of a maximal dipolar potential ${\cal E}_c$. It is natural to speculate that  ${\cal E}_c$ increases with the coupling and diverges at strong coupling. It would be interesting to understand the mechanism behind this effect. Another possibility is that in a concrete realisation like the ABJM theory, the gravity computation is different (due to the existence of non-vanishing scalar fields) and it also 
gives rise to a maximum electric field. We plan to study this question in the near future.

\section*{Acknowledgements}
J. E. Santos would like to thank Benson Way and Gary Horowitz  for helpful discussions.  The research leading to these results has received funding from the [European Union] Seventh Framework Programme [FP7-People-2010-IRSES] and [FP7/2007-2013] under grant agreements No 269217, 317089 and No 247252, and from the grant CERN/FIS-NUC/0045/2015. \emph{Centro de F\'isica do Porto} is partially funded by the Foundation for  Science and Technology of Portugal (FCT). L.G. and M.O. are funded by the FCT/IDPASC fellowships SFRH/BD/51983/2012 and PD/BD/113486/2015, respectively. This work was partially undertaken on the COSMOS Shared Memory system at DAMTP, University of Cambridge operated on behalf of the STFC DiRAC HPC Facility. This equipment is funded by BIS National E-infrastructure capital grant ST/J005673/1 and STFC grants ST/H008586/1, ST/K00333X/1.

\appendix

\section{\label{ap:I}Perturbative expansion to order ${\cal E}^3$}
In order to get an analytic understanding of our results for small electric field ${\cal E}$ and also to test our numerical results, we decided to perform a perturbative expansion to ${\cal E}^3$. Higher orders can be readily obtained, but the functions involved in the expansion become increasingly more complicated.

We start by detailing the generic procedure, which should be valid to any order in perturbation theory. Since the stress energy tensor is even in the Maxwell field, the expansion in powers of ${\cal E}$ will take the following schematic form
\begin{equation}
g_{\mu\nu} = \bar{g}_{\mu\nu}+\sum_{j=1}^{+\infty} g^{(2j)}_{\mu\nu}{\cal E}^{2j}\,,\quad \quad\quad 
{\cal A}_{\mu}=\sum_{j=0}^{+\infty} a^{(2j+1)}_{\mu}{\cal E}^{2j+1}\,.
\end{equation}
We are interested in solutions where the only nontrivial component of the Maxwell field is ${\cal A}_t$. Furthermore, for $\bar{g}$ we choose $AdS_4$ written in global coordinates with standard spherical coordinates, \emph{i.e.}
\begin{equation}
\bar{g}_{\mu\nu}  dx^\mu  dx^\nu = -\left(1+\frac{r^2}{l^2}\right)  dt^2+\frac{ dr^2}{1+\frac{r^2}{l^2}}+r^2\left[\frac{ d\chi^2}{1-\chi^2}+(1-\chi^2) d\phi^2\right] ,
\end{equation}
where $\chi\in[-1,1]$ can be related to the standard polar coordinate as $\chi = \cos \theta$, and we reintroduced the $AdS$ length $l$.

At this point, we choose a gauge. A convenient gauge is the so-called \emph{quasi spherical-gauge} \cite{Bartnik:1996js}, in which the full solution, to all orders in $\mathcal{E}$, can be written as
\begin{equation}
 ds^2 = -\left(1+\frac{r^2}{l^2}\right) Q_1(r,\chi) dt^2+Q_2(r,\chi)\,\frac{ dr^2}{1+\frac{r^2}{l^2}}+Q_3(r,\chi)\,r^2\left[\frac{ d\chi^2}{1-\chi^2}+(1-\chi^2) d\phi^2\right] .
\end{equation}
This gauge completely fixes all gauge redundancy if and only if the functions $Q_i$, with $i\in\{1,2,3\}$, depend on $r$ and $\chi$. If the $Q_i$ depend on $r$ only, then one is still able to fix $Q_3=1$. Our generic expansion in $\mathcal{E}$ can now be applied to our specific line elements, and yields
\begin{equation}
Q_i(r,\chi) = 1+\sum_{j=1}^{+\infty} q_i^{(2j)}(r,\chi)\,{\cal E}^{2j}\,,\quad \quad\quad 
{\cal A}_t(r,\chi) = \sum_{j=0}^{+\infty} a^{(2j+1)}_{t}(r,\chi) \,{\cal E}^{2j+1}\,.
\end{equation}

At linear order in ${\cal E}$ one obtains a second order differential equation for $a^{(1)}_t$ that can be readily solved using separation of variables
\begin{equation}
a^{(1)}_t(r,\chi) = \sum_{\ell=0}^{+\infty}a_\ell\,\frac{\Gamma\!\left(\frac{\ell+1}{2}\right)\Gamma\!\left(\frac{\ell+3}{2}\right)}{\Gamma\!\left(\ell+\frac{3}{2}\right)}{}_2F_1\left(\frac{\ell}{2}+\frac{1}{2},\frac{\ell}{2},\ell+\frac{3}{2},-\frac{r^2}{l^2}\right) L_\ell(\chi)\,,
\end{equation}
where $L_\ell(\chi)$ is a Legendre polynomial of degree $\ell$, ${}_2F_1$ is the Gaussian Hypergeometric function and the $a_\ell$ are real numbers that depend on the harmonic number $\ell$ and which fully specify the boundary chemical potential. In particular, the factors of $\Gamma$ ensure that
\begin{equation}
\lim_{r\to+\infty}a^{(1)}_t(r,\chi) = \sum_{\ell=0}^{+\infty}a_\ell L_\ell(\chi)\,.
\end{equation}
In our concrete example, we want $a_1 = 1$ and $a_\ell=0$ for $\ell\neq 1$. This in turn gives the following expression for $a^{(1)}_t$
\begin{equation}
a^{(1)}_t(r,\chi)=\frac{2}{\pi  r^2}\left[\left(l^2+r^2\right) \arctan\left(\frac{r}{l}\right)-l r\right]\chi\,.
\label{eq:linear1ap}
\end{equation}

We can now proceed to second order. Essentially, we want to solve for a metric perturbation sourced by a stress energy tensor generated by (\ref{eq:linear1ap}). In four spacetime dimensions, metric perturbations about spacetimes which have $SO(3)$ symmetry fall within one of two classes: scalar-type gravitational perturbations and vector-type gravitational perturbations \cite{Regge:1957td,Zerilli:1974ai,Moncrief:1974am,Kodama:2003jz}. Within our symmetry class, the quasi spherical gauge kills all vector-type modes, and we are just left with the scalars which makes the problem considerably simpler.

Scalar-type gravitational modes are labelled by spherical harmonics of degree $\tilde{\ell}$. Since at the linear level the gauge field consists of a single harmonic with $\ell=1$, we can use the usual decomposition of the product of spherical harmonics into its sum to conclude that the metric perturbation will admit the following decomposition
\begin{align}
Q_i = \alpha_i(r) L_0(\chi)+\beta_i(r) L_2(\chi)\,,
\end{align}
where we can use the residual gauge freedom of the sector independent of $\chi$ to set $\alpha_3(r)=0$. It is rather trivial to solve these equations subject to normalisability at the conformal boundary and regularity at the centre of $AdS$. For completeness, we present the final expressions below
\begin{subequations}
\begin{multline}
\alpha_0(r) = \frac{l^2}{3 \pi ^2 r^4 \left(l^2+r^2\right)}\Bigg[4 l^4 r^2+3 \left(\pi ^2-4\right) l^2 r^4-8 l r \left(l^4+3 l^2 r^2+3 r^4\right) \arctan\left(\frac{r}{l}\right)+\\ 4 \left(l^6+l^4 r^2-3 l^2 r^4-3 r^6\right) \arctan\left(\frac{r}{l}\right)^2+3 \pi ^2
   r^6\Bigg]\,,
\end{multline}
\begin{equation}
\alpha_1(r) =-\frac{8 l^5}{3 \pi ^2 r^4 \left(l^2+r^2\right)} \left[l \left(l^2+r^2\right) \arctan\left(\frac{r}{l}\right)^2+l r^2-\left(2 l^2 r+r^3\right) \arctan\left(\frac{r}{l}\right)\right]\,,
\qquad\qquad\qquad\ \ 
\end{equation}
\begin{multline}
\beta_1(r) = \frac{l^3}{12 \pi ^2 r^4 \left(l^2+r^2\right)}\Bigg\{-l r^2 \left[\left(9 \pi ^2-8\right) l^2+5 \left(8+3 \pi ^2\right) r^2\right]+32 l \left(l^2+r^2\right)^2 \arctan\left(\frac{r}{l}\right)^2+\\
r \left[\left(9 \pi ^2-40\right) l^4+2 \left(9 \pi ^2-8\right) l^2 r^2+\left(9 \pi ^2-40\right) r^4\right] \arctan\left(\frac{r}{l}\right)\Bigg\}\,,
\end{multline}
\begin{multline}
\beta_2(r) = \frac{l^3}{12 \pi ^2 r^4 \left(l^2+r^2\right)}\Bigg\{l r^2 \left[\left(56+9 \pi ^2\right) l^2+5 \left(8+3 \pi ^2\right) r^2\right]+32 l \left(l^2+r^2\right)^2 \arctan\left(\frac{r}{l}\right)^2-\\
r \left[\left(88+9 \pi ^2\right) l^4+2 \left(56+9 \pi ^2\right) l^2r^2+\left(9 \pi ^2-40\right) r^4\right] \arctan\left(\frac{r}{l}\right)\Bigg\}\,,
\end{multline}
\begin{multline}
\beta_3(r) = -\frac{l^2}{12 \pi ^2 r^4}\Bigg\{r^2 \left[12 \pi ^2 r^2-\left(8+9 \pi ^2\right) l^2\right]+l r \left[\left(9 \pi ^2-8\right) l^2-\left(56+9 \pi ^2\right) r^2\right] \arctan\left(\frac{r}{l}\right)+\\
16 \left(l^4+4 l^2 r^2-3 r^4\right) \arctan\left(\frac{r}{l}\right)^2\Bigg\}\,.
\end{multline}
\end{subequations}

At third order in $\mathcal{E}$ the calculation becomes more complicated. In particular, for $a^{(3)}_t$ one now has the following decomposition
\begin{equation}
a^{(3)}_t = f_1(r) L_1(\chi)+f_3(r) L_3(\chi)\,,
\end{equation}
where normalisability and regularity dictate
\begin{subequations}
\begin{multline}
f_1(r) = \frac{8 l^2}{525 \pi ^3 r^6}\bigg[25 l^6+247 l^4 r^2-425 l^2 r^4+
\\
\qquad \quad
288 r^4 \left(l^2+r^2\right) \log \left(\frac{2 i r}{l+i r}\right)+288 i l r^5-359 r^6\bigg]\arctan\left(\frac{r}{l}\right)^3
\\
+ \frac{l^2}{350 \pi ^3 r^5}\arctan\left(\frac{r}{l}\right)^2 \Bigg\{4608 i r^3 \left(l^2+r^2\right) \text{Li}_2\left(\frac{l-i r}{l+i r}\right)+
\\
l \left[15 \left(7 \pi ^2-8\right) l^4-2 \left(1816+105 \pi ^2\right) l^2 r^2+4608 r^4 \log\left(\frac{2 i r}{l+i r}\right)-\left(5752+315 \pi ^2\right) r^4\right]\Bigg\}+
\\
\frac{l^2}{5250 \pi ^3 r^3}\Bigg\{25 \left(88+63 \pi ^2\right) l^3+8640 r \left[4 l r \text{Li}_3\left(\frac{l-i r}{l+i r}\right)-8 i \left(l^2+r^2\right) \text{Li}_4\left(\frac{l-i r}{l+i r}\right)-9 l r \zeta (3)\right]
\\
+768 i \pi ^4 l^2 r+10 l r^2 \left[13512+\pi ^2 (576 \log 2-193)\right]+768 i \pi ^4 r^3\Bigg\}+
\\
\frac{l^2}{525 \pi ^3 r^4} \arctan\left(\frac{r}{l}\right) \Bigg\{864 r^2 \left[12 \left(l^2+r^2\right) \text{Li}_3\left(\frac{l-i r}{l+i r}\right)+9 \zeta (3) \left(l^2+r^2\right)+8 i l r \text{Li}_2\left(\frac{l-i r}{l+i r}\right)\right]
\\
-15 \left(16+21 \pi ^2\right) l^4-4 l^2 r^2 \left[2510+\pi ^2 (144 \log 2-127)\right]+2 \pi ^2 r^4 (359-288 \log 2)\Bigg\}\,,
\end{multline}
\begin{multline}
f_3(r) = \frac{8 l^2}{525 \pi ^3 r^6}\Bigg[175 l^6+313 l^4 r^2-240 i l^3 r^3+195 l^2 r^4-48 r^2 \left(l^2+r^2\right) \left(5 l^2+r^2\right) \log \left(\frac{2 i r}{l+i r}\right)
\\
-208 i l r^5+89 r^6\Bigg] \arctan\left(\frac{r}{l}\right)^3+\frac{l^2}{700 \pi ^3 r^5}\arctan\left(\frac{r}{l}\right)^2 \Bigg\{l \Bigg[35 \left(9 \pi ^2-136\right) l^4+6 \left(984+245 \pi ^2\right) l^2 r^2-\\
512 \left(15 l^2 r^2+13 r^4\right) \log \left(\frac{2 i r}{l+i r}\right)+\left(3944+1155 \pi^2\right) r^4\Bigg]-1536 i r \left(l^2+r^2\right) \left(5 l^2+r^2\right) \text{Li}_2\left(\frac{l-i r}{l+i r}\right)\Bigg\}+
\\
\frac{l^2}{525 \pi ^3 r^4} \arctan\left(\frac{r}{l}\right) \Bigg\{5 l^4 \left[2016+\pi ^2 (515+96 \log 2)\right]-384 i \left(15 l^3 r+13 l r^3\right) \text{Li}_2\left(\frac{l-i r}{l+i r}\right)-\\
432 \left(l^2+r^2\right) \left(5 l^2+r^2\right)\left[3 \zeta (3)+4 \text{Li}_3\left(\frac{l-i r}{l+i r}\right)\right]+3 l^2 r^2 \left[560+\pi ^2 (589+192 \log 2)\right]+2 \pi ^2 r^4 (48 \log 2-89)\Bigg\}
\\
+\frac{l^2}{31500 \pi ^3 r^4} \Bigg\{2880 \bigg[24 i \left(l^2+r^2\right)\left(5 l^2+r^2\right) \text{Li}_4\left(\frac{l-i r}{l+i r}\right)-4 \left(15 l^3 r+13 l r^3\right) \text{Li}_3\left(\frac{l-i r}{l+i r}\right)+
\\
9 l r \zeta (3) \left(15 l^2+13 r^2\right)\bigg]-3840 i \pi ^4 l^4-75 l^3 r \left[6328+\pi ^2 (2249+384 \log 2)\right]
\\
-4608 i \pi ^4 l^2 r^2-20 l r^3 \left[11016+\pi ^2 (5561+1248 \log 2)\right]-768 i \pi ^4 r^4\Bigg\}\,,
\end{multline}
\end{subequations}
where $\text{Li}_i(x)$ is a polylogarithm function of order $i$, and $\zeta(x)$ is the Riemann zeta function. 
At linear order in $\mathcal{E}$ we recover the results of \cite{Herdeiro:2015vaa}.

From our third order expansion one can compute all the quantities presented in the text. For instance, the location, in terms of proper distance along the axis defined by $\theta = 0,\pi$, of point like charged extremal particles in our perturbative backgrounds is given by (in units of $\ell = 1$)
\begin{equation}
\mathcal{P}^\star_{\theta = 0}=\frac{4}{3 \pi }\mathcal{E}+\frac{\mathcal{E}^3}{14175 \pi ^3}\left[77760 \zeta (3)+33400-9 \pi ^2 (979+1152 \log 2)\right]+\mathcal{O}(\mathcal{E}^5)\,.
\end{equation}

\bibliographystyle{utphys}
\bibliography{./PBHinAdS}

\begin{thebibliography}{10}

\bibitem{Ernst}
F.~J. {Ernst}.
\newblock {A new family of solutions of the Einstein field equations.}
\newblock {\em Journal of Mathematical Physics}, 18:233--234, 1977.

\bibitem{Fefferman:2007rka}
Charles Fefferman and C.~Robin Graham.
\newblock {The ambient metric}.
\newblock 2007.

\bibitem{MiguelMSc}
Miguel Oliveira.
\newblock {Spectral Methods in General Relativity: Polarized Black Holes in AdS
  Space}.
\newblock {\em Porto University M.Sc thesis}, October, 2013.

\bibitem{MiguelEssay}
Miguel Oliveira.
\newblock {Polarized AdS Black Holes}.
\newblock {\em Essay presented at MAP-fis meeting (Minho, Aveiro, and Porto
  Universities joint Ph.D program)}, September, 2014.

\bibitem{Dias:2015nua}
Oscar J.~C. Dias, Jorge~E. Santos, and Benson Way.
\newblock {Numerical Methods for Finding Stationary Gravitational Solutions}.
\newblock 2015.

\bibitem{gr-qc/9209012}
J.~David Brown and James~W. York, Jr.
\newblock {Quasilocal energy and conserved charges derived from the
  gravitational action}.
\newblock {\em Phys. Rev.}, D47:1407--1419, 1993.

\bibitem{Skenderis:2002wp}
Kostas Skenderis.
\newblock {Lecture notes on holographic renormalization}.
\newblock {\em Class. Quant. Grav.}, 19:5849--5876, 2002.

\bibitem{hep-th/9902121}
Vijay Balasubramanian and Per Kraus.
\newblock {A Stress tensor for Anti-de Sitter gravity}.
\newblock {\em Commun. Math. Phys.}, 208:413--428, 1999.

\bibitem{deHaro:2000xn}
Sebastian de~Haro, Sergey~N. Solodukhin, and Kostas Skenderis.
\newblock {Holographic reconstruction of space-time and renormalization in the
  AdS / CFT correspondence}.
\newblock {\em Commun. Math. Phys.}, 217:595--622, 2001.

\bibitem{Frolov:2006yb}
Valeri~P. Frolov.
\newblock {Embedding of the Kerr-Newman black hole surface in Euclidean space}.
\newblock {\em Phys. Rev.}, D73:064021, 2006.

\bibitem{arxiv:0804.1832}
David Kastor.
\newblock {Komar Integrals in Higher (and Lower) Derivative Gravity}.
\newblock {\em Class. Quant. Grav.}, 25:175007, 2008.

\bibitem{hep-th/9902170}
Andrew Chamblin, Roberto Emparan, Clifford~V. Johnson, and Robert~C. Myers.
\newblock {Charged AdS black holes and catastrophic holography}.
\newblock {\em Phys. Rev.}, D60:064018, 1999.

\bibitem{hep-th/9904197}
Andrew Chamblin, Roberto Emparan, Clifford~V. Johnson, and Robert~C. Myers.
\newblock {Holography, thermodynamics and fluctuations of charged AdS black
  holes}.
\newblock {\em Phys. Rev.}, D60:104026, 1999.

\bibitem{hep-th/9908001}
Bo~Sundborg.
\newblock {The Hagedorn transition, deconfinement and N=4 SYM theory}.
\newblock {\em Nucl. Phys.}, B573:349--363, 2000.

\bibitem{hep-th/0110196}
Alexander~M. Polyakov.
\newblock {Gauge fields and space-time}.
\newblock {\em Int. J. Mod. Phys.}, A17S1:119--136, 2002.

\bibitem{hep-th/0310285}
Ofer Aharony, Joseph Marsano, Shiraz Minwalla, Kyriakos Papadodimas, and Mark
  Van~Raamsdonk.
\newblock {The Hagedorn - deconfinement phase transition in weakly coupled
  large N gauge theories}.
\newblock {\em Adv. Theor. Math. Phys.}, 8:603--696, 2004.
\newblock [,161(2003)].

\bibitem{arXiv:0806.1218}
Ofer Aharony, Oren Bergman, Daniel~Louis Jafferis, and Juan Maldacena.
\newblock {N=6 superconformal Chern-Simons-matter theories, M2-branes and their
  gravity duals}.
\newblock {\em JHEP}, 10:091, 2008.

\bibitem{Bartnik:1996js}
Robert Bartnik.
\newblock {Einstein equations in the null quasispherical gauge}.
\newblock {\em Class. Quant. Grav.}, 14:2185--2194, 1997.

\bibitem{Regge:1957td}
Tullio Regge and John~A. Wheeler.
\newblock {Stability of a Schwarzschild singularity}.
\newblock {\em Phys. Rev.}, 108:1063--1069, 1957.

\bibitem{Zerilli:1974ai}
F.~J. Zerilli.
\newblock {Perturbation analysis for gravitational and electromagnetic
  radiation in a reissner-nordstroem geometry}.
\newblock {\em Phys. Rev.}, D9:860--868, 1974.

\bibitem{Moncrief:1974am}
V.~Moncrief.
\newblock {Gravitational perturbations of spherically symmetric systems. I. The
  exterior problem.}
\newblock {\em Annals Phys.}, 88:323--342, 1974.

\bibitem{Kodama:2003jz}
Hideo Kodama and Akihiro Ishibashi.
\newblock {A Master equation for gravitational perturbations of maximally
  symmetric black holes in higher dimensions}.
\newblock {\em Prog. Theor. Phys.}, 110:701--722, 2003.

\bibitem{Herdeiro:2015vaa}
Carlos Herdeiro and Eugen Radu.
\newblock {Anti-de-Sitter regular electric multipoles: Towards
  EinsteinÐMaxwell-AdS solitons}.
\newblock {\em Phys. Lett.}, B749:393--398, 2015.

\end{thebibliography}
\end{document}